\documentclass[aps,prx,reprint,notitlepage]{revtex4-2}
\usepackage{graphicx} 
\usepackage{physics}
\usepackage{tikz}
\usepackage[dvipsnames]{xcolor}
\usepackage{soul}
\newcommand{\mctwo}{Department of Microtechnology and Nanoscience -- MC2,
Chalmers University of Technology, SE-41296 Gothenburg, Sweden}

\newcommand{\cell}{{\mbox{\scriptsize cell}}}

\begin{document}

\title{Nature of frontier quasi-particle states in nitrogen-base systems}

\author{Raul Quintero-Monsebaiz}
\email[]{raulq@chalmers.se}
\affiliation{\mctwo}
\author{Per Hyldgaard}
\email[]{hyldgaar@chalmers.se}
\affiliation{\mctwo}
\author{Elsebeth Schr{\"o}der}
\email[]{schroder@chalmers.se}
\affiliation{\mctwo}

\date{October 7, 2025}

\begin{abstract}
Understanding photophysical properties of DNA is important: It can help us elucidate and probe the impact of charges and free radicals in the
cellular environment. 
For example, a photoemission at a given nucleobase means that we both charge it and place an electron right next to a neighboring part of the genetic code. 
Inverse photoemission means that we trap a free electron (at some
empty state or resonance), and instead emit a low-energy photon. 
This may reduce the damage if it happens at an already charged base, but it can cause extra damage if it arises somewhere else. 
Predicting the nature of sudden optically-driven excitations, termed quasi-particles (QPs), help us detail 
interactions and possibly control the damage that might follow. Also, these QPs contain information on the larger DNA assembly because they reflect the fingerprints of nucleobase polarity, the hydrogen bonding in Watson-Crick pairs, and the van der Waals (vdW) interactions in the Watson-Crick-pair stacking that makes up the genome.   In this study, we utilize the recently developed (optimally tuned) range-separated hybrid vdW density functional, AHBR-mRSH* [JPCM {\bf 37}, 211501 (2025)] to analyze the electron-attached and ionized QP states of these DNA components, with a particular focus on dipole- and multipole-trapped empty  states (bound or resonances). We also evaluate critical properties such as dipole and quadrupole moments, QP HOMO-LUMO energy gaps, and transition-dipole moments.  Finally, we classify the Watson-Crick stacked dimers based on their QP nature. This classification provides the foundation for proposing a model of DNA reactivity and photo-physical activity.
\end{abstract}

\maketitle
\section{Introduction}

Noncovalent interactions, including dispersive forces, are essential for the sequence-dependent stability of the genome \cite{hunter1993}. 
Deoxyribonucleic acid (DNA) contains the essential genetic instructions for the development, functioning, and reproduction of all known living organisms. 
The nitrogenous bases that constitute the core of the DNA structure are adenine (A), guanine (G), cytosine (C), and thymine (T) \cite{WATSON1953}. 
Combinations that match either A-and-T or C-and-G bases in planar structures are so-called Watson-Crick (WC) pairs that have a central role in biology \cite{cleaves2011}. 
Within DNA, each of the nitrogenous bases sits covalently bonded to a ribose sugar, which is connected to phosphates, forming the sugar-phosphate backbone. 
Two such strings of base-sequences can, in turn, combine to a double-helix form when arranged so that every base on one helix string match up with the uniquely specified WC match on the other string. 
In effect, DNA can be seen as storing the genetic code with redundancy via the unique sequencing of essentially parallel steps 
formed by WC pairs stacked (at relative twists  $\sim 36^\circ$) 
with separations of about 3.3 {\AA} \cite{cooper08p1304,licothlula09,shukla2022}. 
While hydrogen bonds contribute to the formation of planar WC pairs, van der Waals (vdW) (here defined as dispersive) forces combine with electrostatics to dominate in setting the DNA cohesion energy: It is an extended system formed by ring-shaped organic components \cite{sauers1995,RanPRB16}. 
The WC-step separations equal those found in layered systems \cite{rydberg2003,berland2015} and vdW forces play a central role of setting, for example, the detailed structure in the (twisted) WC stacking in DNA \cite{cooper08p1304,licothlula09,jhunjhunwala2021,love2024}. 
A consistent vdW inclusion in first-principles theory characterizations secures accuracy also for individual nitrogen bases \cite{le12p424210,shukla2022,schroder2025}.

Quasi-particles (QPs) define the (linear-response) electronic structure and are thus a key concept of the many-body perturbation theory (MBPT) of both light-matter interactions and of molecules and materials \cite{hedin1965,AuJoWi00}. 
QP predictions obtained in an exact MBPT study are believed to capture all or close to all
details of the photon-driven charging events under two assumptions: 
1) that the reactions happen suddenly and all excitation-relevant details of electron relaxations are completed before ensuing atomic
relaxations occur, and 
2) that we have consistently captured all relevant interactions to the environment. 
For such ideal-MBPT QPs, their energies reflect that of the photon that is either absorbed or emitted in connection with sudden electron releases or attachments, and the spatial structure (so-called Dyson orbital) defines the matrix elements in a Fermi's golden rule description. 
Approximations to such QP orbitals are a starting point for crafting exchange-correlation (XC) functionals \cite{helujpc1971} 
and hence
for making density functional theory (DFT) better at material descriptions. More directly, they provide insight on elementary steps in radiation damage.

In living organisms, DNA and genome stability depends not only on internal noncovalent interactions but also on its interactions within various cellular environments.  When these interactions are induced by radiation, they can lead to mutations, aging, and various diseases \cite{teoule1987,wallace1994,fuciarelli1995}. 
Notably, recent studies have shown that low-energy electrons, produced as secondary products from the interaction between high-energy radiation and water, play a crucial role in DNA damage \cite{barrios2002,zheng2006,gao2021}. 
These phenomena have driven extensive theoretical \cite{baccarelli2011,nikjoo1997,xifeng2006} and experimental research \cite{bachorz2008,alizadeh2013,pater2014,schlatholter2024}. 
Electron attachment (also called electron affinity, in both cases abbreviated as EA) to DNA has been identified as one of the key steps in radiation-induced DNA damage \cite{rak2015,kumar2024}.
Also, it is known that high-energy ionizing radiation induces DNA breaks by generating primary and secondary species, ultimately leading to cell death \cite{ward1988,alizadeh2015}. 
Due to its destructive capacity, such high-energy and ionizing radiation is widely used in radiotherapy treatments \cite{lomax2013}. 
For such events the ionization potential (IP) of DNA defines the energy required to remove an electron from, e.g., a base.

However, since DNA is a macromolecule, studying its EA and IP QPs remains challenging. The simplest models of genetic material are the nucleobases, and MBPT  or other \textit{ab-initio} studies of the EA and IP QPs \cite{nguyen2016,macnaughton2005,ribot2015,dutta2015} are crucial for understanding radiation-damage pathways. It has been shown \cite{qian2011,cooper2023} that the first ionization (highest occupied molecular orbital, HOMO) state is primarily contained on the $\pi$-lobes of the aromatic-type rings of the bases.
In contrast, the first unoccupied (lowest unoccupied molecular orbital, LUMO) state is contained further from the nuclear region at the end of the base dipole vector.
These weakly confined orbitals or resonances \cite{fennimore2018}, are identified as dipole- or (when relevant) multi-pole trapped states, 
having a small electron-nucleus binding energy but also small kinetic energies.

Understanding photo-driven 
charging in WC pairs is perhaps even more important: While their selectivity 
is a fundament of having genome information, the fact that reversible
non-covalent interactions provide 
this selectivity means that Nature can 
also preserve it upon genome transcription \cite{Schrodinger1944}. Here, theoretical studies of the EA in the C-G pair using MP2/6-31++G** calculations have shown that the dipole-trapped state is stable \cite{smets2001}. 
Later, density functional theory (DFT) studies using the B3LYP functional predicted the dipole-trapped anion of A-T \cite{richardson2002,jihad2000}. 
More recently, electron-attachment equation-of-motion coupled-cluster calculations with domain-based local pair natural orbitals and single and double excitations (EA-EOM-DLPNO-CCSD) using a pVTZ + 5s5p4d basis set revealed that the dipole-trapped anion lies near C as a large lobe in C-G pairs and near T in A-T pairs, while the valence-bound anions are located on C and T, respectively \cite{tripathi2019}. 
Studies of IPs in both gas and aqueous phases have also been conducted for C-G base pairs using the B3LYP-D3/6–311+G** method \cite{BeckeIII,LYP,grimme3,uddin2024}.

Furthermore, when WC pairs are stacked as in DNA, they form a system of four nucleobases, commonly referred to as a base-pair step or Watson-Crick dimer (WCd). These systems can adopt 10 distinct configurations and this 
set is relevant as a minimal model for the genetic encoding.
The stacking energies of WCd systems have been benchmarked using various computational approaches, including dispersion-corrected density functional approximations \cite{cablo2015}, by use of vdW density functionals (vdW-DFs) \cite{cooper08p1304,shukla2022,frostenson2022,schroder2025}, by variants of MP2 and MP2-F12 \cite{metha2022}, several high-order symmetry-adapted perturbation theory (SAPT) methods \cite{parker2013}, as well as selected low-cost and semi-empirical methods \cite{kruse2019,ken2020}. 

Nevertheless, their electron-attached and ionized states remain underexplored. 
In fact, of the above-listed papers, it is only our introduction of an optimally tuned (OT) range-separated 
hybrid \cite{HSE03,KuKrRev2008,smiga2020} (RSH) vdW-DF form \cite{schroder2025}, termed
AHBR-mRSH*, that explicitly computed and discussed excited resonances and states across the nitrogen bases and WC pairs, 
while also predicting QP energies for three WCds. 
For individual nucleobases there are both GW characterizations  \cite{qian2011} and
Koopmans-DFT studies \cite{li2017,nguyen2016,nguyen2018,colonna2022} of the HOMO- and LUMO-type QPs, 
and our AHBR-mRSH* results are consistent with those predictions. Our AHBR-mRSH* method simply adapts planewave DFT to extract QPs 
predictions subject to an OT criterion \cite{kronik2012,schroder2025} 
that is shown to also give de facto compliance with a charge-transfer condition for molecules \cite{PeLe83,KuKrRev2008}. The QP 
studies listed in the previous paragraphs
rely instead on large Gaussian basis sets, with the exception of Ref.\ \cite{jihad2000}, and do not include vdW interactions. However, vdW forces are clearly relevant, at least for understanding bound states 
in WCds: Use of the vdW-inclusive AHBR-mRSH gives a dramatic improvement of 
the description of the inter-base and especially WCd binding, compared with its precursors \cite{kronik2012,schroder2025}.

In this paper, we perform robust and consistent predictions of QP states, dipole moments, and transition dipole moments for nucleobases and WC pairs, as well as across the full set of WCds. We again employ the above-mentioned AHBR-mRSH* with OT choice for the 
RSH screening crossover $\gamma=0.166$
that we asserted for nitrogen-base systems in Ref.\ \onlinecite{schroder2025}.  
The success of AHBR-mRHS* as a predictor of both total-energy differences and for the frontier QP has been tested in various benchmark sets, showing its predictor capability \cite{schroder2025}.

The outline of the paper is as follows: Sections II and III present the methodology employed to obtain the results discussed in Section IV. We include a discussion of selecting atomic geometries for the nitrogen-base components and of the nomenclature used for the WCd  systems.  Section  IV also  
provides a visualization for the frontier QPs and interpret their nature as correlated with their electrostatics moments while  suggesting a WCd classification scheme. 
Section V presents our conclusions. Finally, the paper also contains an appendix on estimating
the quadrupole moments of the WCds.

\section{Modeling and theory}

\begin{figure}[tb]
    \centering
    \includegraphics[width=8cm]{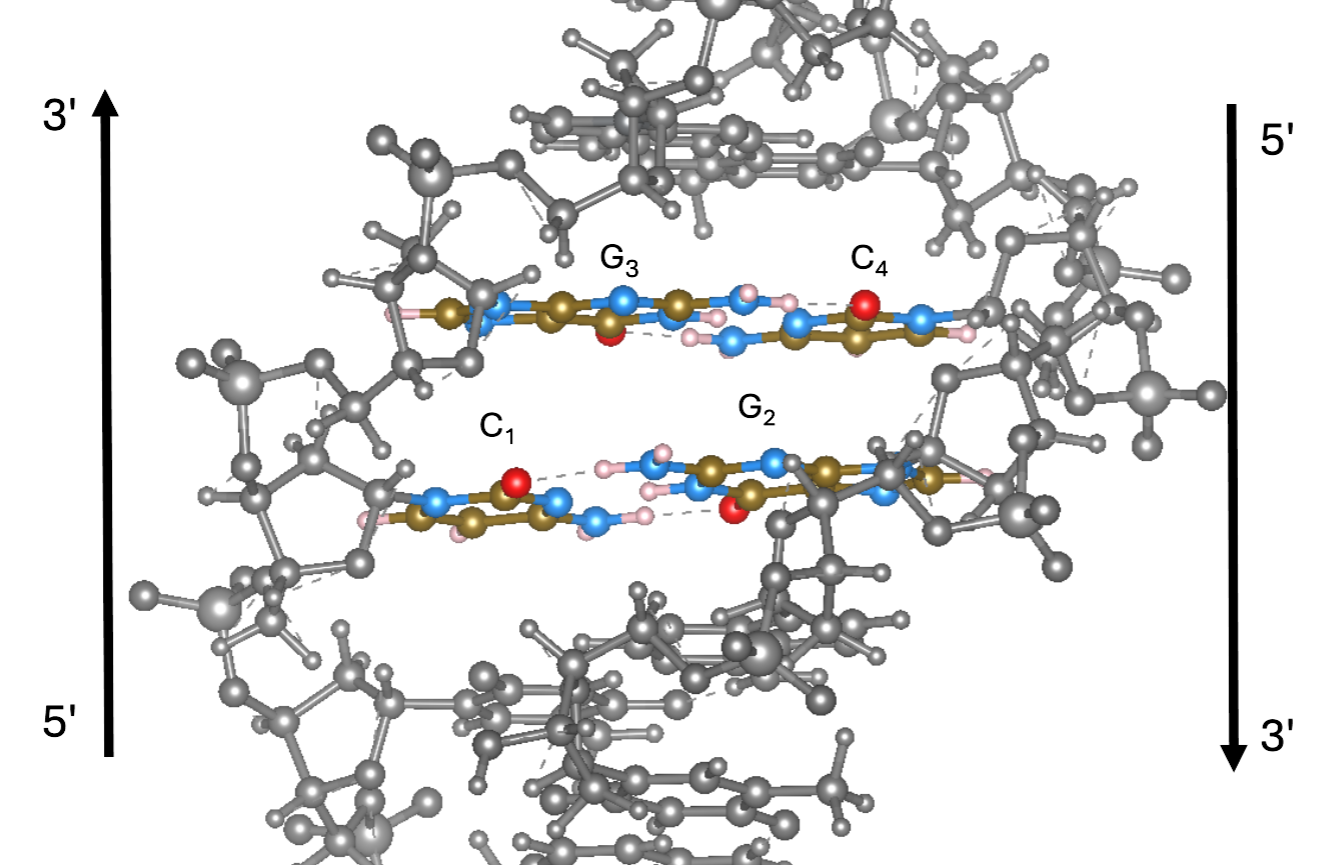}
    \caption{Four nucleobases (colored atoms) comprise the Watson-Crick pair dimer (WCd) attached to the DNA sugar phosphate backbone (gray atoms), viewed from the minor groove. 
    With a nucleobase $L_1$ ($L_2$) in the lower left (right) corner of the base pair step and with
    nucleobase $L_3$ ($L_4$) in the upper left (right) corner, we label the step $L_1$p$L_3$.
    This is here exemplified in the figure with C$_1$, G$_2$, G$_3$, and C$_4$, yielding CpG.
    In the colored sections, red/brown/blue/white balls represent oxygen/carbon/nitrogen/hydrogen atoms, respectively.
}
    \label{fig:parallelepiped}
\end{figure}

We use AHBR-mRSH* \cite{schroder2025} to predict and analyze the set of QP frontier states of
nucleobases (A, G, C, and T), of the 
two WC pairs (A-T and C-G), and of the ten distinct WCds (denoted ApA, GpC, ApT, TpA, CpG, GpC, ApG, ApC, TpC, and TpG) that can be found as the inside of a two-step segment of DNA. 
These structures are investigated without inclusion of the backbone helix strings, at representative geometries, Ref.\ \onlinecite{kruse2019}.

Fig.\ \ref{fig:parallelepiped}
illustrates an example organization 
of four nucleobases (with nucleobase atoms highlighted in color) in a DNA segment and gives the logic of the nomenclature for the set of ten WCds. At every DNA step, a base is matched into a WC pair. The details of the backbone strings (outermost atoms in gray) are set by the sequence of groups, and
those run in opposite order on the pair of helix strings, as illustrated by the pair of outside arrows. Within we find the WC-pairs, linked by hydrogen bonds (dashed line) as well as vdW forces. 
We also find  WCds, formed when the WC pairs are stacked at a $\sim 36^\circ$ relative twist \cite{cooper08p1304}. 
The nomenclature is explained in the caption of Figure \ref{fig:parallelepiped}. 
We exhaust the 
qualitative variation in DNA building blocks up to two-step segments by including the ten above-listed WCds.
Still, by focusing our studies of DNA building blocks with fixed coordinates listed in Ref.\ \onlinecite{kruse2019}, we can at most claim to be exploring the nature of QPs in such motifs.

Computing the IP and EA QPs of WC pairs and WCds  requires that long-range interactions, such as vdW (dispersion) forces, must be included and simultaneously accurate for both atomic-structure and QP prediction.  
DFT within the Kohn-Sham (KS) scheme \cite{hohenberg1964,kohn1965} can 
use, for example, (nonhybrid)
versions of the vdW-DF method \cite{rydberg2000,rydberg2003,dion2004,thonhauser2007,berland2014,hyldgaard2020,berland2014a} to 
obtain often very accurate predictions of forces, the electron-density variation, as well
as a set of (KS) orbital energies at a very low computational cost 
for any of the here-investigated 
DNA subsystems. 
One popular such density functional approximation (DFA) is the rev-vdW-DF2 
\cite{lee10p081101,hamada2014,callsen2015}, also denoted vdW-DF2-b86r (abbreviated B86R).
The resulting KS-HOMO and KS-LUMO
do represent the chemical potentials for electron removal and addition, respectively. Such KS-HOMO and KS-LUMO states also provide a fair approximation of the gaps for transitions \cite{levy1984,almbladh1984,chong2002,weitao2012,gritsenko2003,onida2002}
but they are not, themselves, valid
as QP predictions. For example, the 
KS-LUMO generally fails to capture the dipole-trapped nature of the actual QP-LUMO in nucleobases \cite{fennimore2018,colonna2022,cococcioni2005,schroder2025}.

Problems with a direct application of KS-DFT for QP predictions arise largely because (nonhybrid) DFAs exhibit a convex curvature when plotting the total energy against the number of electrons, whereas it should ideally be piecewise linear \cite{perdew1982,LevyGKS,schroder2025}. 
Several strategies have emerged to obtain QP energies within DFT  \cite{li2017, Ma2016,schubert2023,gennaro2022,nguyen2018,davo2010,linscott2023,liechtenstein1995, dudarev1998,cococcioni2005}. Use of RSHs with full asymptotic screening 
of Fock-exchange contributions (like the HSE \cite{HSE03,HSE06} or RSH vdW-DFs \cite{shukla_2022a,shukla2022,frostenson2022}) often permit us to ameliorate localization errors in the resulting predictions by generalized KS (gKS) DFT \cite{LevyGKS}. 
This is a benefit that is partly correlated with restoration of piecewise linearity in the resulting gKS DFA. The RSH vdW-DF2-ahbr (abbreviated AHBR and based on a so-called analytical-exchange-hole modeling of the exchange in B86R) 
is proving itself useful for general structure and energy predictions, from solids to molecules \cite{shukla2022,Cu2O2}.

For use in small molecule problems, however, it is motivated to assume that an unscreened Fock exchange description characterizes the interactions between an electron and its associated hole \cite{gulu76,lape77} at large relative separations \cite{KuKrRev2008}.
This observation leads to the formulation of a class of closely-related molecular-type RSHs, denoted AHBR-mRSH($\gamma$) \cite{schroder2025}, where a parameter $\gamma$ is used to control the assumed cross-over between the assumptions 
for screening of exchange at short ranges and at asymptotic electron-hole separations \cite{HSE03,kronik2012}. Like AHBR, 
the AHBR-mRSH($\gamma$) class of RSH vdW-DFs  give total-energy descriptions that are accurate for molecules for a broad range of $\gamma$ values \cite{schroder2025}. We may therefore, as in the correspondingly defined HSE-based OT-RSHs \cite{kronik2012,OTRSHalga,OTRSHgap}
use an OT scheme to set $\gamma$
so that AHBR-mRSH*=AHBR-mRSH($\gamma^*$)
is at the same time also a QP predictor. We find that using the same
$\gamma^*$ value (obtained by looking at the adiabatic ionization of T), gives a resulting AHBR-mRSH* form that ensures strict piecewise-linearity behaviors for multiple bases and highly accurate predictions of frontier QPs \cite{schroder2025}.

Below we use the AHBR-mRSH* to now predict 
and interpret the nature of frontier QP levels across the broader set of DNA building blocks. 
We keep AHBR-mRSH* unchanged from that used in Ref.\ \onlinecite{schroder2025}, i.e., with the OT step set by the nature of ionization in individual nitrogen bases.

\section{Computational details}

We employ Quantum ESPRESSO (QE) \cite{giannozzi2009,giannozzi2017,linACE,PaoloElStruct1} with our in-house implementation of AHBR-mRSH* and with the B86R \cite{lee10p081101,hamada2014} DFA as a KS-DFT for comparisons. 
We use optimized norm-conserving Vanderbilt (ONCV) pseudopotentials \cite{hamann2013,sg15} with a planewave cutoff set to 160 Ry.   
To suppress the electrostatic interactions between periodic cells we place the molecules in a 30 {\AA} cubic cell and use the Makov-Payne correction \cite{markov1995}, with the k-point-sampling limited to the $\Gamma$-point.

\begin{figure}[tb]
    \centering
        \includegraphics[width=8cm]{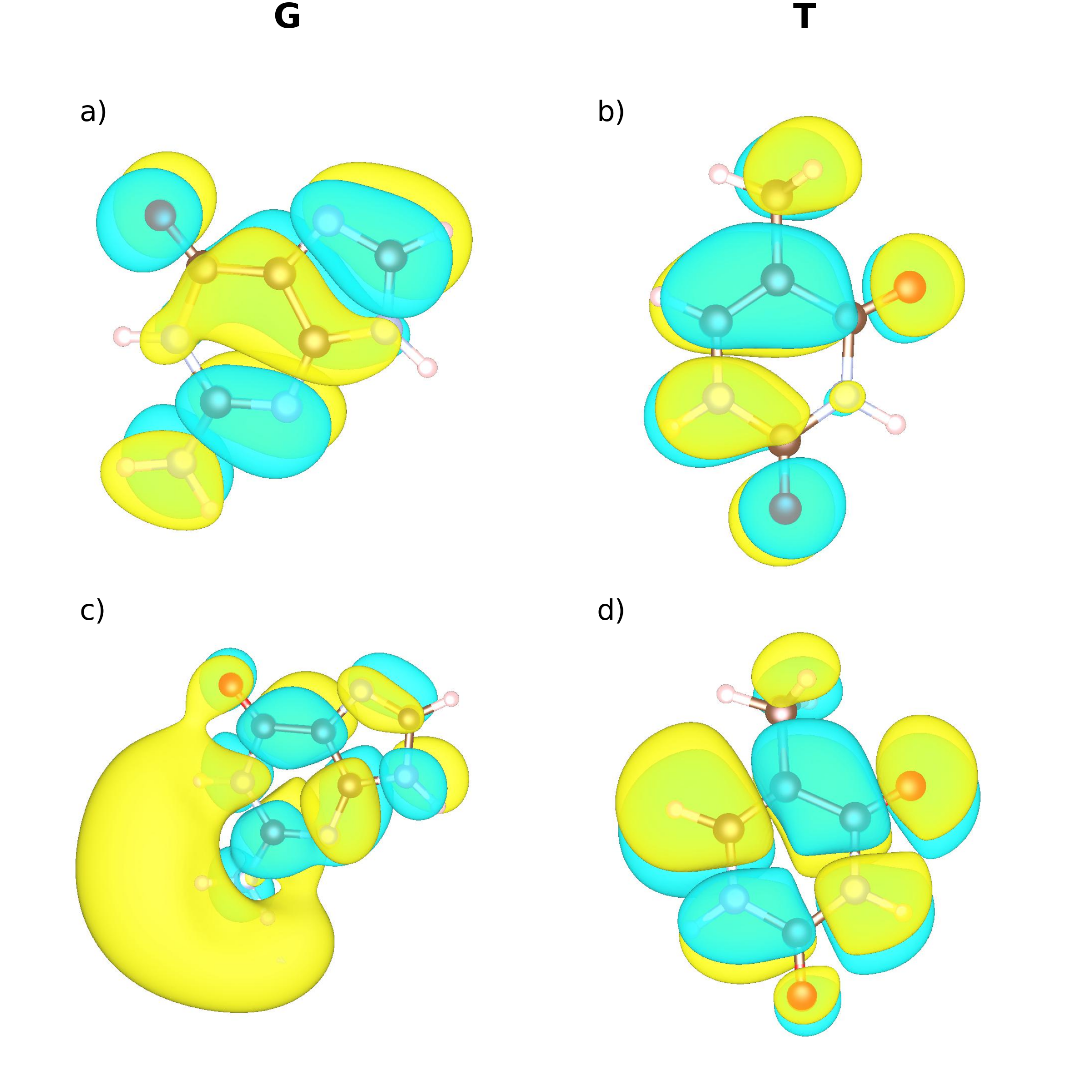} 
    \caption{B86R based KS-HOMO/KS-LUMO orbital densities $|\psi_i(\mathbf{r})|^2$: 
    (a) KS-HOMO of G,  (b) KS-HOMO of T, (c) KS-LUMO of G and (d) KS-LUMO of T. 
Yellow and blue isosurfaces correspond to different signs on the  
wavefunctions $\psi(\mathbf{r})$
(that are evaluated at $k=0$ and have no complex phase).  
      }
    \label{fig:GT_b86r}
\end{figure}

The dipole moments are obtained directly from the DFT calculations, while the wave function squared 
$|\psi_i(\mathbf{r})|^2$ for the orbitals, also called the orbital density, is calculated using the post-processing program \texttt{pp.x} of the Quantum Espresso suite. Here, $i$ corresponds to the KS-HOMO/KS-LUMO (in B86R calculations) or the QP-HOMO/QP-LUMO (in AHBR-mRSH* calculations). 
For the orbital densities 
we plot isosurfaces at
approximately $7\times 10^{-5}$ $a_{0}^{-3}$
where $a_0$ denotes the 
Bohr length scale.

The transition dipole moment is computed using an in-house program, and is defined as
\begin{equation}
     M_{LH}=
    \mel{\psi_{L}}{ \hat{\boldsymbol{r}}}{\psi_{H}}=\int_{\cell} \psi_{L}^{*} \qty(\mathbf{r}) \hat{\boldsymbol{r}} \psi_{H}\qty(\mathbf{r}) d \mathbf{r}\,.
\end{equation}
Here, $\psi_{L}^{*} \qty(\mathbf{r})$ is the complex conjugate of the LUMO state, 
$\hat{\boldsymbol{r}}$ is the position operator and $\psi_{H}\qty(\mathbf{r})$ is the HOMO state. 
To perform this integral numerically we employ the trapezoid quadrature in an homogeneous grid of 420 points per axis. 

We compute and map the
nonlocal-correlation or vdW contributions to the inter-WC-pair cohesion in the WCds as follows.
We use the post-DFT-processing \textsc{ppACF} code \cite{JiScHy18a,hyldgaard2020,MOFdobpdc} to extract 
local energy densities,
\begin{equation}
   e_c^{\rm nl}(\mathbf{r})  \equiv  \frac{n(\mathbf{r})}{2} \int 
    \Phi[n](\mathbf{r}, \mathbf{r}^{\prime}) \, n\left(\mathbf{r}^{\prime}\right) 
    \, 
    d{\mathbf{r}^{\prime}} \, ,     
\end{equation}
where $\Phi[n](\mathbf{r},\mathbf{r'})$ denotes the kernel of the $E_c^{\rm nl}$ term in AHBR \cite{dion2004,thonhauser2007,hyldgaard2020}.
We do this for both the entire
WCd system and for the two constituent WC pairs (kept at the same coordinates as in WCd). 
We use the 
standard postprocessing tool of QE for defining the net
cube file reflecting the spatial variation in the difference,
$\Delta e_c^{\rm nl}(\mathbf{r})$,
in $E_c^{\rm nl}$ local binding energy that  results with the WCd formation, as also done in Ref.\ \onlinecite{hyldgaard2020}.

\section{Results and discussion}

The KS-DFT orbitals differ from our QP characterization
of the nitrogen bases and the range of investigated base systems.
Figure \ref{fig:GT_b86r} shows our results for the KS-HOMO and -LUMO orbital densities, $|\psi_i|^2$, as computed for the G nitrogen base, using B86R as the DFA. In such KS-DFT descriptions the (KS-)LUMOs are generally valence-bound states (VBS), which is also what we find for all KS-LUMOs of the bases.

\begin{figure*}
    \centering
        \includegraphics[width=17cm]{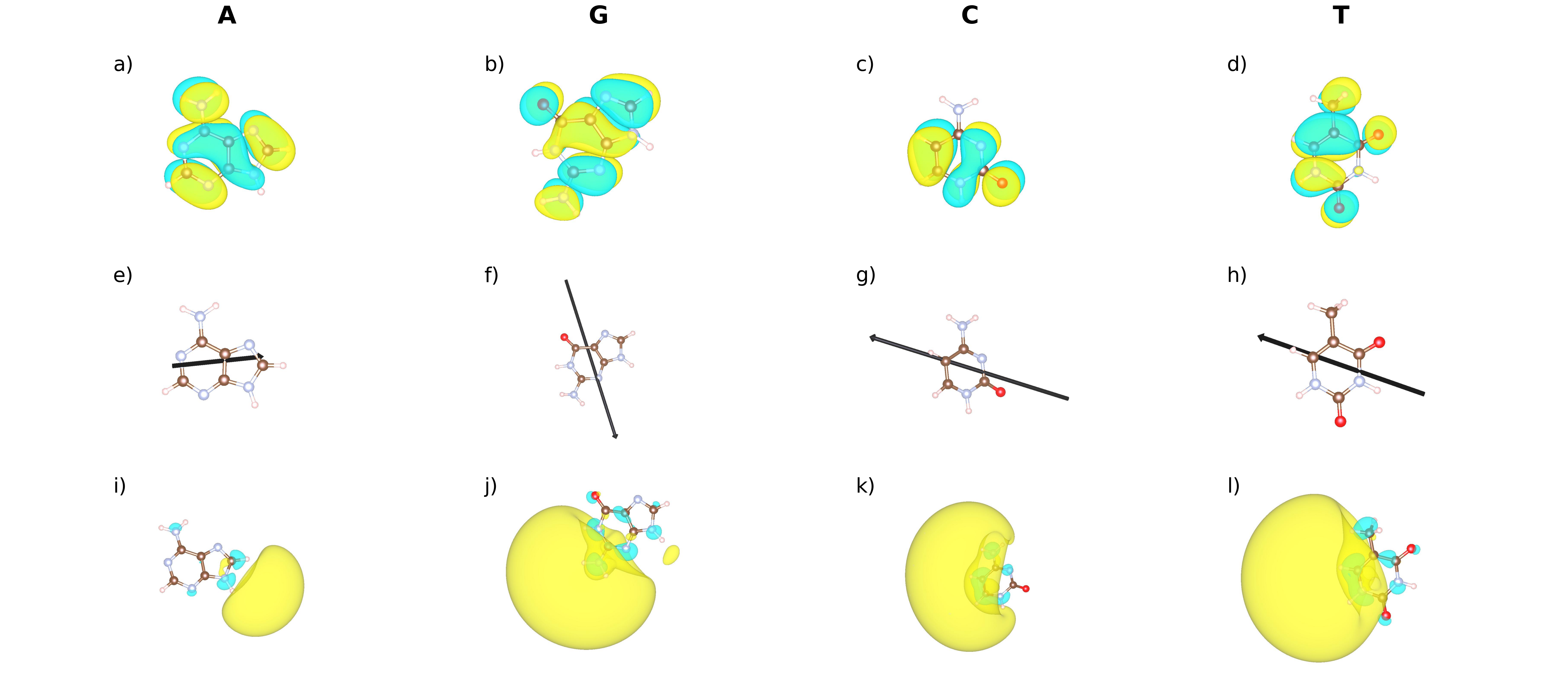} 
    \caption{AHBR-mRSH* based frontier orbital densities and dipole moments for the nucleobases. 
    The first/second/third row shows QP-HOMOs/dipole moment/QP-LUMOs. Atom color coding as in Figure 1.
     }
    \label{fig:nuc_ahbrmrsh}
\end{figure*}

Meanwhile, Figs.\  \ref{fig:tpcdipecnl}-\ref{fig:class3} 
identify key mechanisms, vdW forces and multipole trapping, that 
help define the QPs and show the spatial variation 
in the range of QP-HOMOs and QP-LUMOs. We predict these QPs by
using (instead) AHBR-mRSH* for the range of nitrogen-base systems.
Table \ref{tab:OpticalPropPredict} summarizes this variation
in terms of 
two key QP descriptors of a possible optical transition 
at the fundamental gap, namely the
QP-HOMO/QP-LUMO energy separation $\Delta_{\rm LH}$, and associated 
dipole-transition matrix elements $|M_{\rm LH}|^2$. 
We select these as descriptors of the QP variation between systems, because
they both enter in a lowest-level, frozen-coordinate Fermi-golden-rule estimate \cite{PrendergastGalli}
\begin{equation}
\Gamma_{{\rm H} \to {\rm L}} \propto |M_{\rm LH}|^2
\delta(\hbar \omega - \Delta_{\rm LH} - E_b) 
\label{eq:FermiGR}
\end{equation}
of exciton absorption at radiation frequency
$\omega$. Completing even this initial exciton-absorption 
modeling, for example, by estimating the variations in the 
exciton-binding energies $E_b$, is beyond the present scope.
Still the AHBR-mRSH* predictions of $\Delta_{\rm LH}$ and $|M_{\rm LH}|^2$ values 
give an impression of the impact of QP variations on the photo-physical properties of  DNA building blocks. 

Importantly, we note that our results for the orbital densities
of the QP-LUMOs, Figs.\ \ref{fig:nuc_ahbrmrsh}-\ref{fig:class3},
are markedly different from the KS-LUMO approximations provided as standard KS-DFT, e.g., as provided by a B86R study. The QP orbitals are not VBS but have the nature of dipole- or multipole-trapped states (as we 
shall document and explain below). The QP-LUMOs are therefore instead concentrated beyond the range of atoms of the set of base systems. This is true whether they are computed as bound orbitals or merely (unbound) resonances in our AHBR-mRSH* characterizations.

\subsection{Electrostatic and vdW nature of QP predictions}

\begin{table*}[tb]
\caption{\label{tab:OpticalPropPredict} Summary of our
AHBR-mRSH* characterization of the QP nature in the
set of investigated base systems. 
This overview is given in terms of the QP-HOMO/QP-LUMO gaps, $\Delta_{\rm LH}$, the
frontier-level QP dipole transition moment matrix elements $\abs{M_{\rm LH}}^{2}$, and calculated 
dipoles $d$. Also shown, for comparison, are dipole $d_{\Sigma}$ and quadrupole $\mathbf{Q}$ moments of the WCd estimated from the DFT-calculated A-T and C-G dipoles, as described in the Appendix. The moments are listed 
with respect to a coordinate system that has the $z$-axis aligned with the B-DNA center axis and the origin midway between the two WC pairs. With this choice, all entries of $\mathbf{Q}$ vanish except $Q_{xz}=Q_{zx}$ and $Q_{yz}=Q_{zy}$. We also list the highest eigenvalue of $\mathbf{Q}$, see also Appendix A.}
\begin{tabular}{l|ccccccc}
System & $\Delta_{\rm LH}$ [eV] & 
$\abs{M_{\rm LH}}^{2}$ [$10^{-3}$ a.u.] & $d$ [D] &
$d_{\Sigma}$ [D]  & $Q_{xz}$ [B] & $Q_{yz}$ [B] & Eigenvalue [B]  \\
\hline
\hline
A & 8.22 & 3.9 & 2.41 & 
\\
T & 8.84 & 0.9 & 4.52 &  
\\
C & 8.45 & 10.8 & 6.80 & 
\\
G & 7.65 & 2.9 & 6.79 & 
\\
\hline
\hline
A-T & 8.05 & 0.8 & 1.73 & 
\\
C-G & 7.24 &   0.003 & 6.03 & 
\\
\hline
\hline
ApA & 7.73 & 0.1  & 2.94 & 3.30& -0.69 &  2.56 & 2.65
\\
CpC & 6.56 &  1.0 & 10.51 & 11.47 & 1.33  & -9.13 & 9.22 
\\
\hline
ApT & 7.80 & 3.1 & 0.18 & 0.16& -8.57 &  0    & 8.57 
\\
TpA & 7.78 & 10.5 & 1.37 & 1.91& 7.19  &  0    & 7.19
\\
GpC & 7.20 & 5.6 & 4.97 & 5.33 & -26.77  &  0    & 26.77
\\
CpG & 7.08 & 0.045 & 1.65 & 2.04& 29.42 &  0    &  29.42
\\
\hline
ApG & 6.99 & 0.6 & 7.03 & 7.73& 10.43  & 2.73 & 10.78
\\
ApC & 7.07 & 0.5 & 4.30 & 4.59& -17.67  & -6.39 & 18.79 
\\
TpC & 6.89 & 1.0 & 6.52 & 7.07 & -9.79  & -8.96 &  13.27
\\
TpG & 7.10 &  0.045 & 4.47 & 4.91& 18.31 & 0.17 & 18.31
\\
\end{tabular}
\end{table*}

\begin{figure}[tb]
    \centering
        \includegraphics[width=7.5cm]{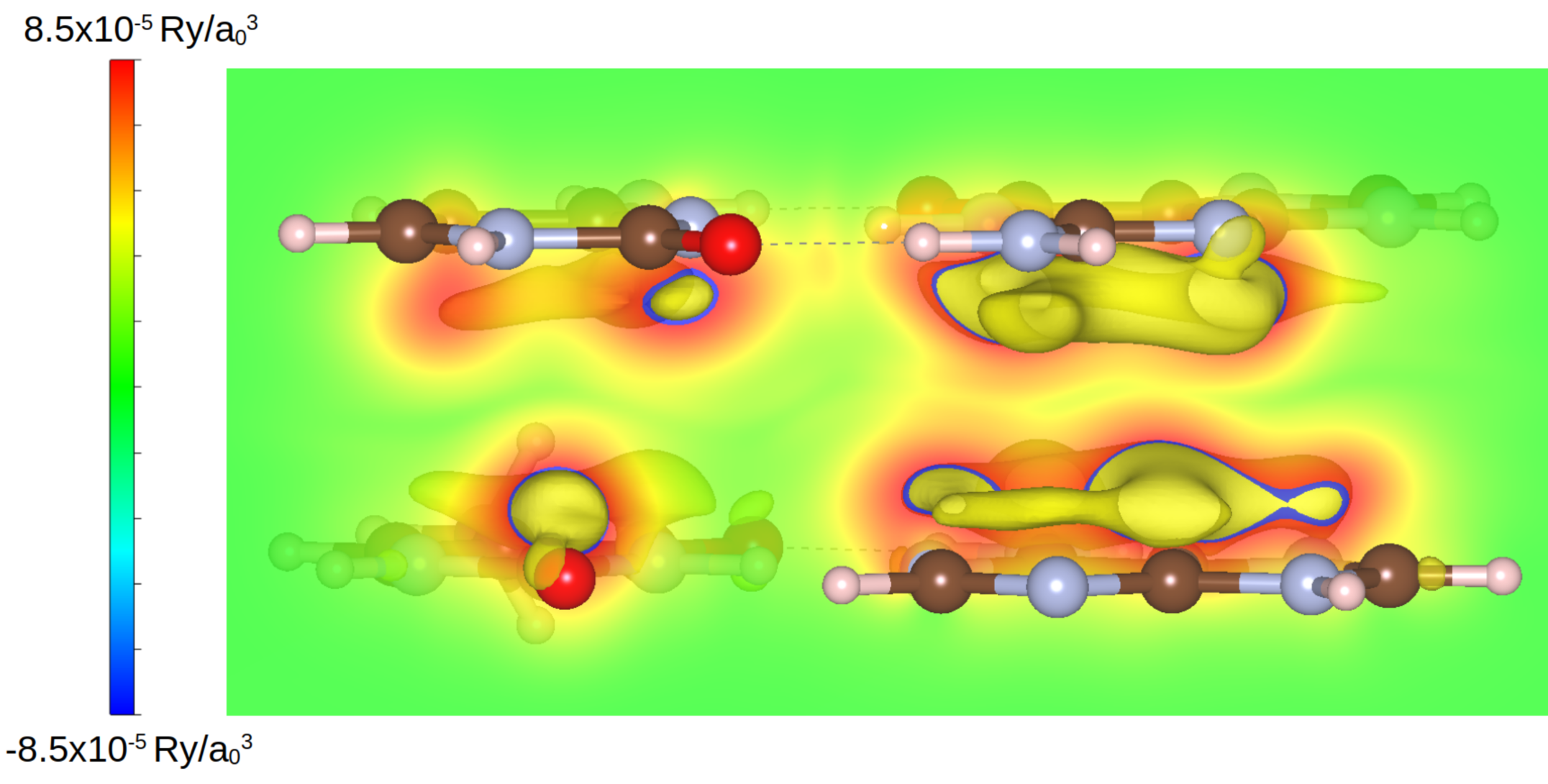}\\[1em] 
         \includegraphics[width=6.5cm]{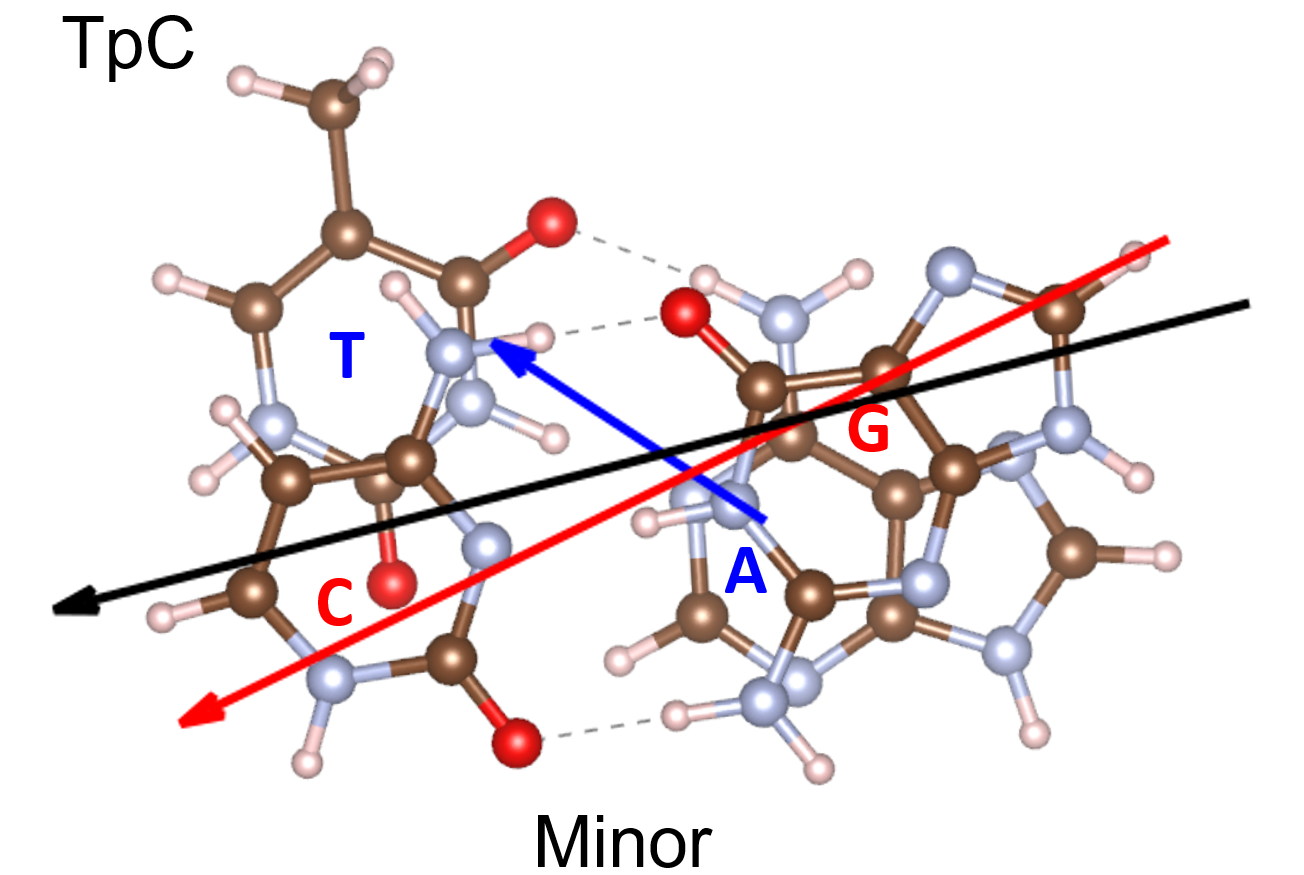}
    \caption{
    \textit{Top panel:} Spatial mapping of the binding energy-density arising from the nonlocal-correlation term of the
    AHBR-mRSH*  DFA for the TpC system, i.e., a ($36^\circ$ twisted) stacked WCd comprising an A-T and an C-G WC pair. 
    For non-covalently bonded systems like ours, this variation is completely set by the concentration of vdW interactions between different electron density regions, as tracked by the yellow contour (set at $-8.8\times 10^{-5}$ Ry/$a_0^3$). 
    The panel also shows binding-energy density variation in a cut through the TpC stack, with color coding revealed by the left-most
    bar. 
    {\it Bottom panel:} Atomic structure of the TpC stack 
    shown together with the DFT result (black arrow) for the net
    dipole that it produces; Atom color coding as in Figure 1. Also shown (to a common scale) are our DFT results for dipoles for the isolated A-T and C-G WC pairs forming components of this TpC stack, blue and red arrows, respectively (and again computed in DFT). Since WCds are held by vdW binding,
    our dipole results for those two WC-pairs
    suffice to get a reasonable estimate of 
    the dipoles and quadrupoles of any of the WCds, see text and appendix.}
    \label{fig:tpcdipecnl}
\end{figure}

The top panel of Fig.\ \ref{fig:tpcdipecnl} reports 
a spatial mapping, $\Delta e_c^{\rm nl}(\mathbf{r})$, of 
the inter-WC-pair binding contribution from nonlocal 
correlations (vdW forces) in the assembly of the TpC 
system (stacking a C-G WC pair on top of an A-T WC pair). 
The plot details the vdW attraction, partly compensated 
by kinetic-energy repulsion and exchange effects
that are ubiquitous in and essential
for an accurate determination of the
cohesion in such base-pair stacks 
\cite{shukla2022,frostenson2022}.
We compute the $\Delta e_c^{\rm nl}(\mathbf{r})$ variation 
by postprocessing directly from the AHBR-mRSH* DFT study that reveals the details of TpC QP nature, below. 

The yellow iso-contour identifies a nonlocal binding-energy 
concentration of $-8.8\times 10^{-5}$ Ry/$a_0^3$. The panel also uses a color coding 
(with values given in the left-most bar) to trace the
range of binding-energy contributions that arise in 
a two-dimensional cut made through the center of the
TpC geometry, approximately what can be seen as the 
shared center of a B-DNA system containing the TpC step or stack. As in
our previous such characterizations of
layered systems
\cite{JiScHy18b,hyldgaard2020,MOFdobpdc,Cu2O2}, we find that the predicted inter-WC-pair 
vdW attraction [identified by large negative values of $\Delta e_c^{\rm nl}(\mathbf{r})$] are centered away from both the planes of 
atoms and away from the mid-separation regions 
\cite{JiScHy18a,hyldgaard2020,frostenson2022,MOFdobpdc,Cu2O2}.

The delocalized nature of the vdW attraction
[the spread in $\Delta e_c^{\rm nl}(\mathbf{r})$] 
that provides WCd cohesion is expected to have ramification for the nature of, at least, the frontier QPs. 
We note that frontier-QP
orbitals extend furthest into the inter-WC-pair regions, where the vdW attraction emerges. 
This suggests that a fully consistent prediction 
of the QP should be based on a vdW-inclusive approach, such as AHBR-mRSH*, for the base systems. 
A future study contrasting the performance of this new (OT) RSH vdW-DF with the corresponding PBE-based OTRSH description to identify vdW fingerprints in the frontier QPs would also be motivated. 
Here we simply note that use of the AHBR-mRSH* leads to at least an order of magnitude accuracy gains
for predictions of WCds, compared to the PBE-based OTRSH \cite{schroder2025}.

The bottom panel of Fig.\ \ref{fig:tpcdipecnl} 
shows the atomic structure of the TpC stacked system, the extent and direction of the net dipole (black arrow) that we can extract directly for our DFT-based QP description. 
The panel shows the TpC stack from a perspective above the bilayer and closest to the C-G WC pair, so that the minor groove of the corresponding DNA segment would be located at the bottom of the figure. 

The bottom panel of Fig.\ \ref{fig:tpcdipecnl} furthermore shows the DFT-computed dipole for two WC pairs (blue and red arrows). 
Using the components of the WC-pair 
dipoles, we can extract good estimates of the WCd dipole as well as for the electrostatic quadrupole moment that we can expect are formed in any given WCd.
 
Table \ref{tab:OpticalPropPredict} summarizes 
our DFT results for length $d$ of the dipole $\vec{d}$ across the range of investigated base systems. 
The table furthermore summarizes estimates, denoted $d_{\Sigma}$, for the length of WCd dipole vectors
as evaluated by first extracting the DFT-computed dipole vector $\vec{d}$ for each of the two WC pairs and then combining those, subject to the assumptions that: 
A) The vectors of the per-WC-dipoles both reside in the plane of the WC pair perpendicular to a shared center line of the B-DNA segment, 
B) They are both centered at this assumed B-DNA center but offset by inter-WC-pair separation $\Delta= 3.3$ {\AA} while twisted by 
36$^\circ$ \cite{cooper08p1304}, and 
that C) We can treat the combined WCd as a sum of the WC-pair charge distributions, i.e., proceed
by simple vector addition to determine 
$d_{\Sigma}$. This latter assumption is motivated by the observations that vdW attraction provides 
the WCd cohesion.
From the DFT calculations of the WC pair dipoles we also find assumptions A) and B) to be fulfilled, Table \ref{tab:OpticalPropPredict}. 

A key observation from the two dipole columns of Table \ref{tab:OpticalPropPredict}
is that the simple dipole estimate $d_{\Sigma}$ indeed aligns with the DFT computed dipole length $d$. That is, 
the numerical results provide a validation
of assumption C). This finding implies, in turn, that we can use the same point-charge model to also extract realistic estimates for all relevant components (and eigenvalues) of the quadrupole moments of the WCds, see Appendix A. Our set of quadrupole results for the set of WCds are listed in the three rightmost columns of Table \ref{tab:OpticalPropPredict}.

Access to simple multipole modeling for the set of nitrogen-base systems is important for our interpretation of the QP-LUMO nature, below. We find that the large electrostatic
moments correlate with
the here-documented systematic tendency for converting the VBS of the KS-DFT description
into the multipole trapping nature of the
QP-LUMOs, i.e., states or resonances that have low electrostatic binding energy but also low kinetic energies. An electron attachment that is represented by this type of QP LUMO, implies that the extra electron sits on the edge of the molecular systems, away from the electrons that originally resided on the occupied orbitals of the neutral base systems.

This overall interpretation of the QP-LUMO nature is, in turn, important for the trust that we put in the following set of detailed QP-LUMO characterizations. 
Like KS-DFT, and like in even a formally exact GW$\Gamma$ determination \cite{hedin1965}
of QPs, our AHBR-mRSH* descriptions are subject to the Born-Oppenheimer approximation: We are implicitly keeping the atomic positions frozen, focusing on capturing as much of the electronic effects as possible in the QP approximation. However, ensuing atomic relaxations will generally also impact 
(vertical-excitation) electron-attachment events, causing deformations. 
If, indeed, the actual frozen-atom QP 
predictions were of the VBS type, the extra electron
would be sitting in the middle of the original
electrons and we must then expect the subsequently atomic relaxations to be large and thus also of larger importance in such events. 
Because AHBR-mRSH* instead predicts multipole-trapped QP-LUMOs, we conversely expect a smaller role from atomic relaxations. This holds, at least, when we (as here) consider photo-physical effects on small nitrogen-base systems in the absence of the DNA backbones. 

\subsection{QP wavefunctions of nucleobases}

For individual nucleobases, two types of anions can be observed: dipole- (or quadrupole-) trapped anions and valence-bound anions or VBS (similar to those shown in Figure \ref{fig:GT_b86r}). 
The QP-LUMO outcomes depend on the choice of methodology and we explicitly focus on the actual QP-LUMO or 
extend the characterizations also to LUMO+1 states, 
Refs.\  \cite{qian2011,nguyen2016,schroder2025}.

One or more unoccupied (LUMO-type) VBS can be captured using conventional KS-DFT methods, as illustrated  also in Figure~\ref{fig:GT_b86r}, where the B86R functional is used. 
There, the orbital densities of the KS-HOMO for T and G are displayed in the top panels, while the KS-LUMO states are shown in the bottom panels. 
The figure shows that the KS-HOMO orbitals are located on the molecule, as expected. 
In G, the KS-LUMO exhibits a slight dipole-trapped character, with parts of the orbital away from the nuclear positions. 
This behavior is not observed for the KS-LUMO of T.

Figure~\ref{fig:nuc_ahbrmrsh} presents our AHBR-mRSH* results of frontier QPs and dipoles for the four individual nucleobases. The first row of panels shows the QP-HOMO states, the second row displays the molecular geometry and dipole moments, and the last row presents the spatial variation in the 
(lowest-unoccupied, i.e., the actual) QP-LUMO states. 
Figure \ref{fig:nuc_ahbrmrsh} shows 
that AHBR-mRSH* predicts dipole-trapped states for the QP-LUMO states of T and G, Fig.~\ref{fig:nuc_ahbrmrsh}(f) and (h), 
in contrast to the VBS character predicted 
in B86R, Fig.~\ref{fig:GT_b86r} bottom panels; 
With B86R we get instead KS-LUMOs that are 
mostly located \textit{on} the molecules. 
Qualitatively, we observe that the QP-LUMO lobe of nucleobase A is the smallest, while that of G is the largest, 
Table~\ref{tab:OpticalPropPredict}.

We find that the
above-stated qualitative observations 
of the positions and shapes LUMO lobes 
of the nitrogen bases correlate with 
the directions and sizes of the dipole 
moments, illustrated to a common 
relative scale, by the vectors shown 
in the set of middle panels of Fig.\ \ref{fig:nuc_ahbrmrsh}.
Importantly, all dipole moments (vectors) point towards the QP-LUMO lobe.  For C, G, and T, the dipole vector points from the oxygen atom (or the center of a pair of oxygen atoms in the case of T) to the opposite side of the nucleobase~\cite{Vovusha2018}, where this lobe is located, since the oxygen atom carries a partial negative charge. For nucleobase A, that has no oxygen atoms, there are three nitrogen atoms on one side and two on the side where the lobe is found. 

\begin{figure}[tb]
    \centering
        \includegraphics[width=7cm]{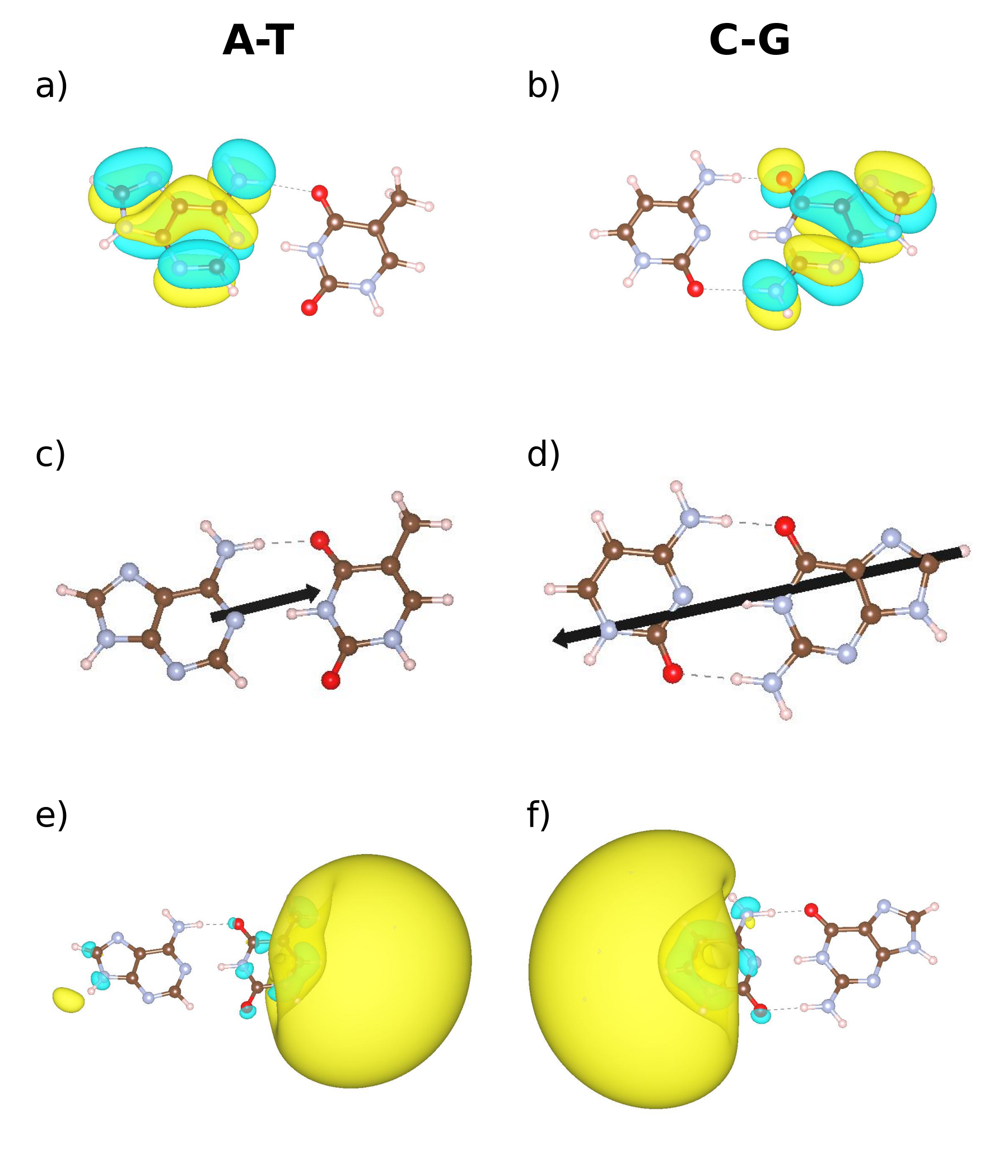} 
    \caption{First/second/third row showing QP-HOMOs/dipole moment/QP-LUMOs for WC pairs, computed in AHBR-mRSH*. 
    Atom color coding as in Figure 1.
    }
    \label{fig:AT_CG_ahbrmrsh}
\end{figure}

In contrast to the LUMO results, the QP-HOMO states do not show any significant changes when compared to the KS-HOMO, Fig.~\ref{fig:GT_b86r} top panels. 
Although the OT choice that converts the generic AHBR-mRSH($\gamma$) functionals into AHBR-mRSH* influences the QP energies it barely affects the binding energies, as demonstrated previously in Ref.\ 
\onlinecite{schroder2025}. It follows that
there is consequently no important impact 
on the occupied-orbital densities $n_{i}(\mathbf{r})$ or the electron density $n \qty (\mathbf{r}) = \sum_{i} n_{i} \qty (\mathbf{r})$.

\begin{table}
\caption{
\label{tab:NucPlusWCpairs} 
Comparison of QP-HOMO-$1$, QP-HOMO, QP-LUMO and QP-LUMO+1 (in eV) for nucleobases, WC pairs, and WCd. The data is computed in AHBR-mRSH*, and compared to experimental values
(listed in the immediately following row, where available).
        }
\begin{ruledtabular}
\begin{tabular}{l|cccc}
       & QP-HOMO-$1$ &  
        QP-HOMO & QP-LUMO & QP-LUMO+1\\
\hline \hline
A & $-9.236$ &   $-8.214$ & 0.007 & 0.121 \\
  & $-9.45^a$ &  $-8.47^a$ & $0.012^d$ & $\sim 0.5^c$ \\
T & $-9.776$ & $-8.910$ & $-0.070$ &  $-0.008$ \\
  & $-10.14^{a}$ & $-9.2^{a}$ & $\sim-0.06^b$ & $\sim 0.4^c$ \\
C & $-9.174$ &  $-8.565$ & $-0.110$ & 0.072 \\
     & $\sim -9.55^{a}$ &  $-8.89^a$ & $-0.23^c$ & $\sim 0.4^c$ \\
G  & $-9.350$ & $-7.816$ & $-0.165$ & 0.050 \\
     & $-9.90^e$ &  $-8.30^e$ & - & - \\
\hline \hline
A-T  & $-8.732$ &   $-8.052$ & 0.007 & 0.106 \\
     &   -  &      -  &   $-0.003^f$   & -  \\
C-G   & $-9.165$ &   $-7.343$ & $-0.108$ & $-0.076$ \\
     &   -  &      -  &   $-0.099^f$   & -  \\
\hline \hline
ApA   & $-7.792$ &   
         $-7.766$ & $-0.032$ & 0.113\\
CpC   & $-7.089$ &   
       $-6.8340$ & $-0.273$ & $-0.089$ \\
\hline 
ApT  &  $-7.9372$ & $-7.8117$  & $-0.0119$ &0.0461 \\
TpA   & $-7.8469$ & $-7.7394$ & 0.0391 & 0.1168 \\
GpC   & $-7.3197$ & $-7.2569$ & $-0.0599$ &  $-0.0128$ \\
CpG   & $-7.2365$ & $-7.0110$ & 0.0666 & 0.1619 \\
\hline
ApG   & $-9.435$ &   
   $-7.095$      & $-0.109$ & 0.082 \\    
ApC  & $-7.9330$ & $-7.1910$ & $-0.1166$ & 0.0697 \\
TpC  & $-7.6957$ & $-7.0701$ & $-0.1795$ & 0.0440 \\
TpG  & $-7.8106$ & $-7.1091$ & $-0.0122$ & 0.1604 \\
\hline
\multicolumn{4}{l}{$^a$Ref.\ \onlinecite{TrScKo06}.}\\
\multicolumn{4}{l}{$^b$Ref.\ \onlinecite{
HeLyCl96}.}\\
\multicolumn{4}{l}{$^c$Collected in Ref.\ \onlinecite{
roca2009}.}\\
\multicolumn{4}{l}{$^d$Ref.\ \onlinecite{
DeAbSc96}.}\\
\multicolumn{4}{l}{$^e$Ref.\ \onlinecite{
DoYoVo78}.}\\
\multicolumn{4}{l}{$^f$Ref.\ \onlinecite{tripathi2019}.}
\end{tabular}
\end{ruledtabular}
\end{table}

\subsection{QP wavefunctions of WC pairs}

Figure \ref{fig:AT_CG_ahbrmrsh} shows our AHBR-mRSH* results for the spatial structure in the HOMO and LUMO QP orbitals of the A-T and C-G pairs. 
As in Fig.\ \ref{fig:nuc_ahbrmrsh}, the middle row of panels shows the atomic structure, along with our AHBR-mRSH* determination of the directions, positions, and relative magnitudes of the dipole moments (with the dipole position at half the length of the vectors). 
The upper row of panels show that in A-T (C-G) the QP-HOMO orbital is located on A (G) with a similar distribution as in the individual nucleobase A (G).

Table \ref{tab:NucPlusWCpairs} list numerical results for the QPs of the nitrogen-base system. 
We find that in the WC pairs the position of the QP-HOMO orbital is determined by the nucleobase with the highest occupied QP orbital energy. 
For example, in  A-T, the QP-HOMO orbital is located on A, with A having higher occupied orbital energy ($-8.2$ eV) than T ($-8.9$ eV).

Panel `e)' of Fig.\ \ref{fig:AT_CG_ahbrmrsh} 
shows that the QP-LUMO of the A-T WC pair is distributed almost entirely in a large lobe connected to T. 
The dipole is pointing towards this lobe that is similar to the QP-LUMO lobe on the individual T  molecule, see Figure~\ref{fig:nuc_ahbrmrsh}. 
Also, panel `f)' of Fig.\ \ref{fig:AT_CG_ahbrmrsh} 
shows that the QP-LUMO of the C-G WC pair is distributed almost entirely in a large lobe connected to C. Again the dipole is pointing towards the QP-LUMO lobe that, in turn, is similar to the QP-LUMO lobe on the individual G base in Figure~\ref{fig:nuc_ahbrmrsh}.

The individual dipole moments of A and T are similar in size and in almost opposite directions, yielding a relative small total dipole moment of A-T WC pair, Table \ref{tab:OpticalPropPredict}.  
In the C-G base pair the QP-LUMO lies outside C, and again the dipoles from the two base pairs do partially compensate each other. However,
the larger size of the two constituent nitrogen-base dipoles makes for a larger dipole in the C-G WC pair than in the A-T WC pair, Table \ref{tab:OpticalPropPredict}, and Appendix A.

\subsection{QP wavefunctions of WCds}

WCd are WC pairs arranged in a stacked configuration and vdW interactions play a key role, as illustrated in
Fig.\ \ref{fig:tpcdipecnl}.
The binding energies of the WCds (also known as base-pair stacking energies) were analyzed in Ref.~\onlinecite{schroder2025} using the AHBR-mRSH* functional, yielding excellent agreement with
quantum chemistry reference calculations \cite{kruse2019}. 
Here we characterize the
effect that the vdW-dominated binding has on the WCd frontier QP states.

\begin{figure}[tb]
    \centering
        \includegraphics[width=7cm]{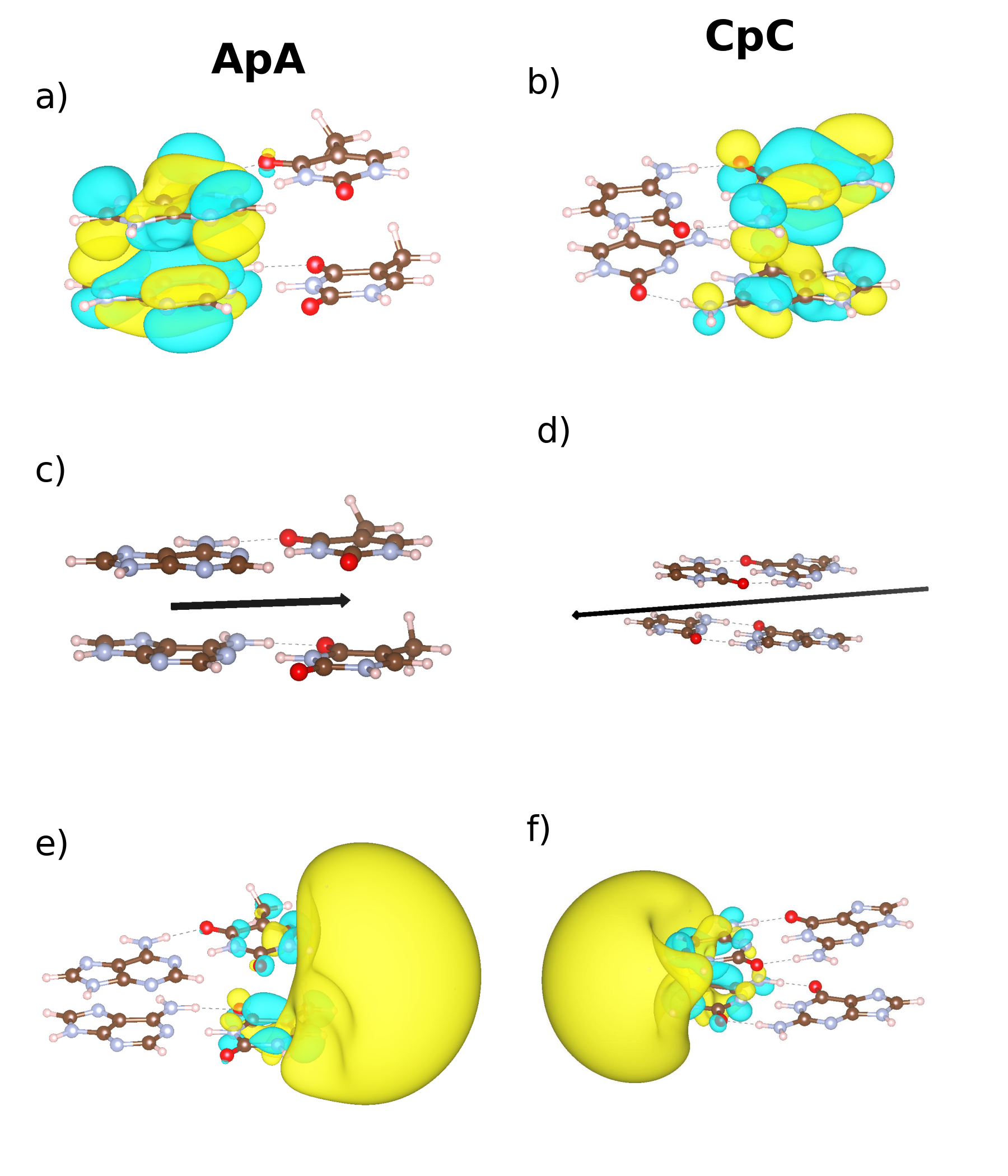} \qquad
    \caption{Class 1 Watson-Crick pair dimers, with first/second/third row showing QP-HOMO/dipole moment/QP-LUMO
     computed in AHBR-mRSH*. The molecules are viewed from the minor groove of DNA.
    }
    \label{fig:class1}
\end{figure}

\begin{figure*}[tb]
    \centering
        \includegraphics[width=17cm]{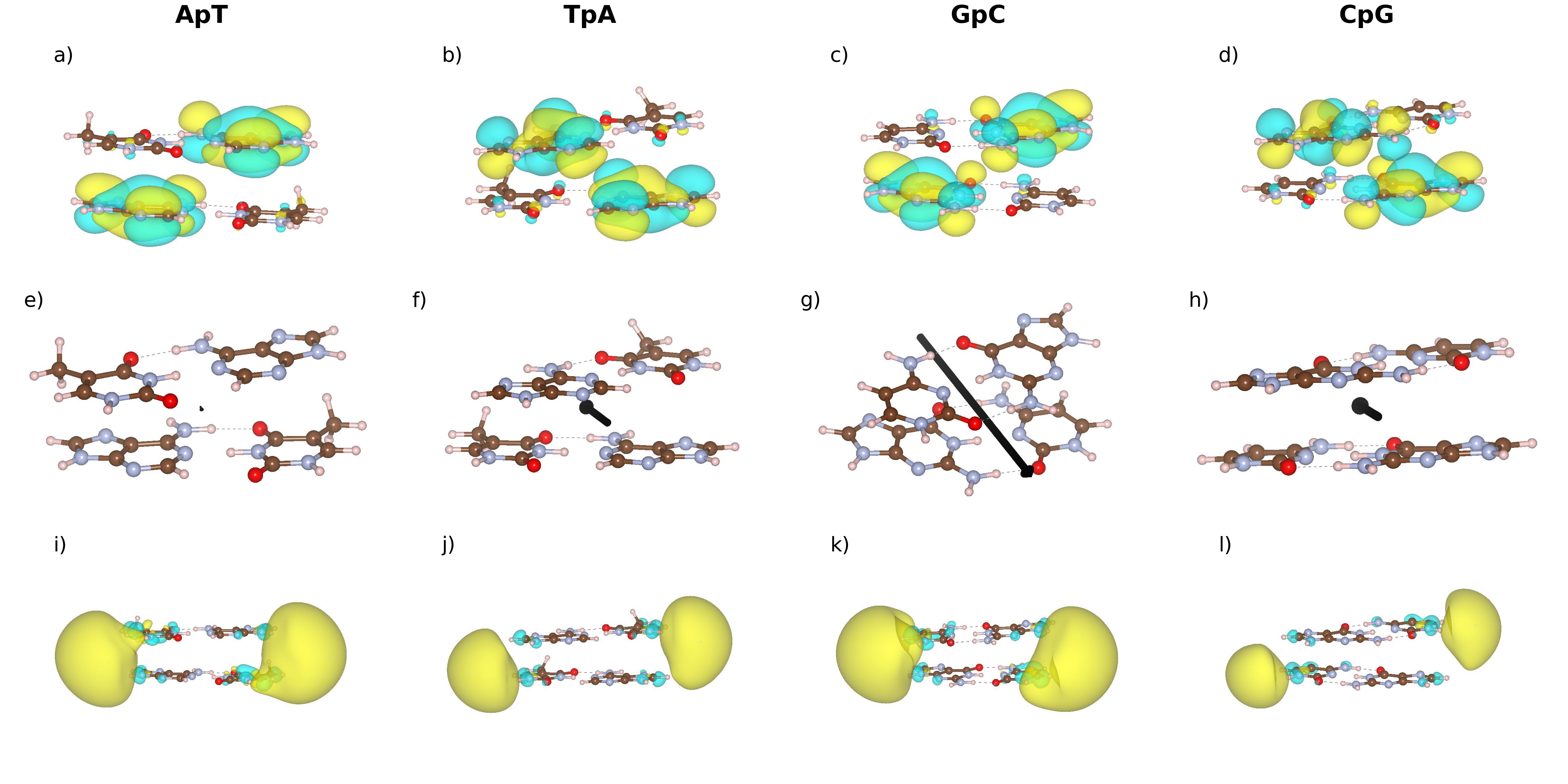} 
    \caption{Calculations with AHBR-mRSH* for Class 2 Watson-Crick pair dimers, panels organized as in Figure \ref{fig:class1}. 
    The dipole moments of ApT and GpC (TpA and CpG) point directly to (from) the minor groove, i.e., out of (into) the paper. 
    Atom color coding as in Figure 1.}
    \label{fig:class2}
\end{figure*}

\begin{figure*}[tb]
    \centering
        \includegraphics[width=17cm]{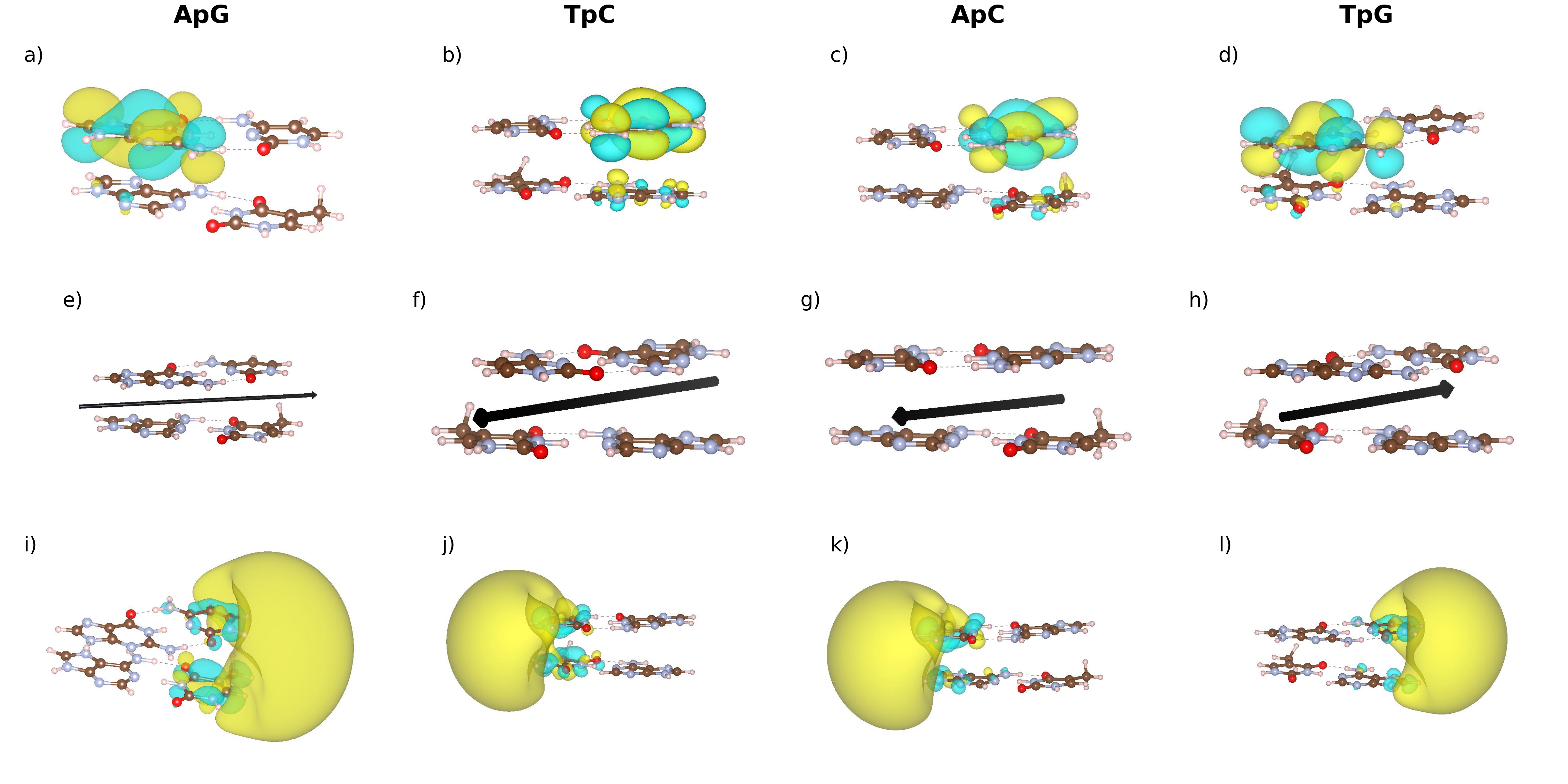} \qquad
    \caption{Calculations with AHBR-mRSH* for Class 3 Watson-Crick pair dimers, panels organized as in Figure \ref{fig:class1}. 
    Atom colors as in Figure 1.
    }
    \label{fig:class3}
\end{figure*}

The WCd form 10 distinct configurations, Fig.\ \ref{fig:parallelepiped}. We categorize these into three different classes, based on the nucleobase content, as follows: 
\begin{description}
    \item[Class 1] ApA and CpC dimers, where bases $L_1$ and $L_3$ are identical (leading to the complementary bases, $L_2$ and $L_4$, also being identical), Figure~\ref{fig:class1}.
    \item[Class 2] Dimers ApT, TpA, GpC, and CpG, where $L_1$ and $L_3$ are complementary, resulting in $L_1 = L_4$ and $L_2 = L_3$, 
    Figure~\ref{fig:class2}.
    \item[Class 3] Dimers that consist of all four nucleobases,  ApG, ApC, TpC, and TpG, Figure~\ref{fig:class3}.
\end{description}

Figure~\ref{fig:class1} shows results for the QP states of Class-1 WCds.
For ApA the QP-HOMO (top panel) is located on the two A bases.
Additionally, we observe orbital density between the two A molecules.
A similar behavior occurs in CpC, where the QP-HOMO is now located on the two C bases. 
Qualitatively, we again observe that the dipole vectors point toward the end of the WCd where QP-LUMO wavefunctions are located, as in the A-T and C-G pairs (Fig.~\ref{fig:AT_CG_ahbrmrsh}), with the dipole magnitude (illustrated by length of vector) being greater in the CpC dimer, see
also Table \ref{tab:OpticalPropPredict} and appendix A.
The QP-LUMO wavefunctions exhibit a  lobe extending outward from the T and C bases in ApA and CpC, respectively. 
Making also a comparison with the WC pairs shown in Fig.~\ref{fig:AT_CG_ahbrmrsh}, we note that the QP states in both QP-HOMO and QP-LUMO states are here partly shared between the two base pairs and that the sharing in more pronounced in the QP-LUMO states.  
This effect is shown in the set of middle panels
and quantified in Table \ref{tab:OpticalPropPredict}. 
This electrostatic trapping of the QP-LUMO 
means that the sudden EA leads to 
potentially forming a more stable dipole-trapped anion, with the extra electron being
pushed outwards towards the region where the WCds connect to the backbone of DNA \cite{zheng2006}.

Figure~\ref{fig:class2} shows frontier QP states and dipole vectors of the WCds in Class 2. 
For the ApT and TpA cases -- and for the GpC and CpG cases -- the two dimers differ by the stacking order relative to the backbone of a B-DNA segment containing such a step. 
The stacking, in turn, causes difference in direction of the twist around the B-DNA center, Fig.\ \ref{fig:parallelepiped}. 
As a result we get that, for example, ApT and TpA differ in the
direction that we must twist 
the A-T WC pair relative to the T-A pair in the WCd to reflect how these stacks are included in 
a corresponding short segment of B-DNA. 

\begin{figure}[tb]
    \centering
        \includegraphics[width=8.5cm]{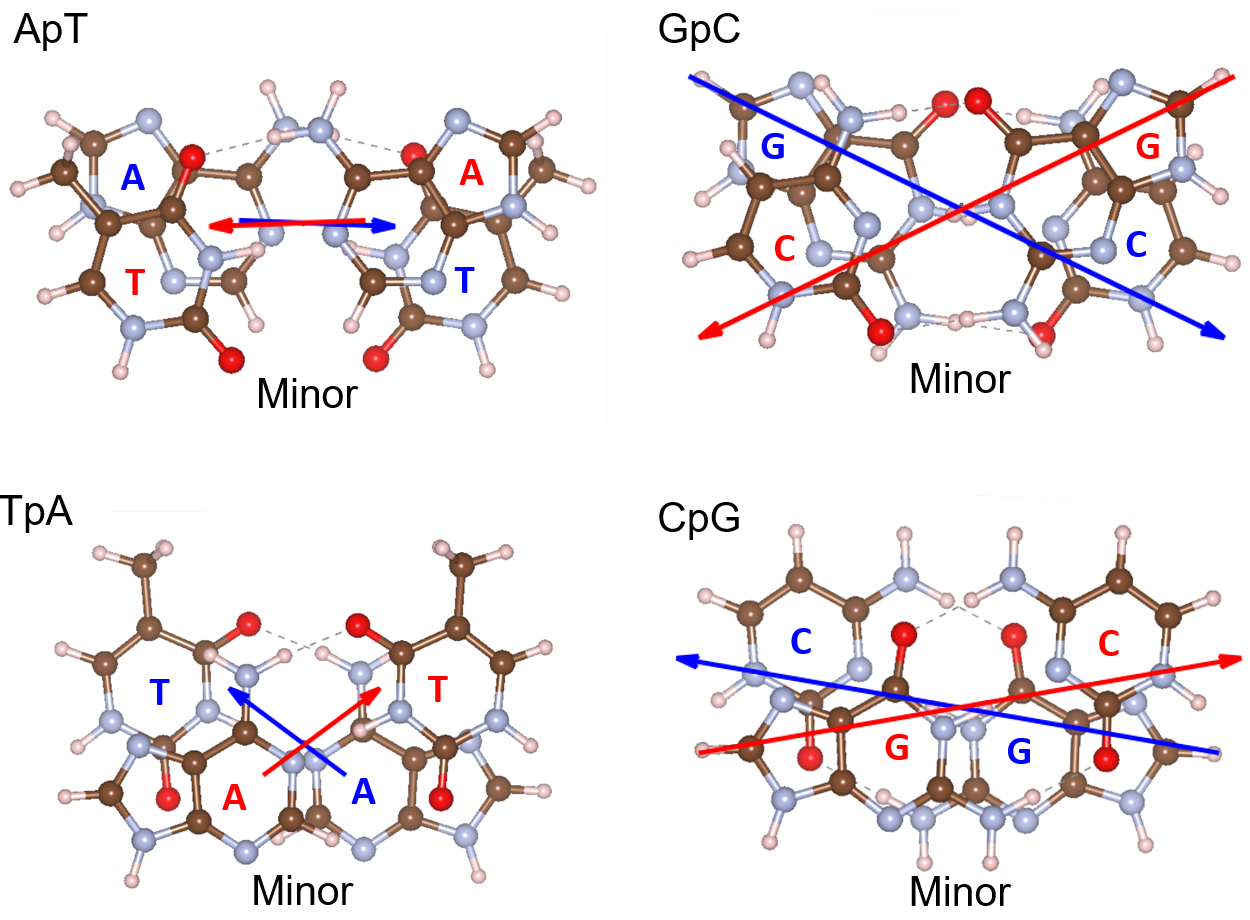} 
    \caption{Relative organization of dipole vectors (red and blue arrows) 
    of constituent WC-pairs in Class-2 WCds. The position of the minor groove is shown. Atom color coding as in Figure 1. Large cancellations arise 
    in the net dipole 
    and motivation and makes it clear that 
    for these systems it is relevant to also consider the impact of a net potential (for electron attachment) produced by  quadrupoles, see Appendix.  
    }
    \label{fig:class2dipoles}
\end{figure}

Figure \ref{fig:class2dipoles} illustrates 
how the twists of the A-T WC pairs lead to differences in the WC dipole orientations, depending on whether the WC pair is the lower or the upper pair
in the WC step.
Blue letters and arrow indicate the molecules of the lower WC pair ($L_1$-$L_2$ in Figure \ref{fig:parallelepiped}) and its dipole, 
while the red letters and arrow indicate the upper WC pair ($L_3$-$L_4$ in Fig.~\ref{fig:parallelepiped}). 
The sum of dipoles $d_\Sigma$ in the WC dimers ApT and TpA are thus different, due to the twist, 
resulting also in differing length in actual dipoles $\vec{d}$ of the
WC dimers. 
Similar effects exist for GpC and CpG, right column of panels in Fig.~\ref{fig:class2dipoles}.
For convenience the position of the minor groove of B-DNA is also shown.
In summary, WCds of Class 2 have dipoles, also shown in middle panels of Fig.\ \ref{fig:class2}, whose magnitude and direction are impacted by the nature of the twists and
set by partially compensating WC-pair contributions.

In ApT and TpA, characterized in the first two columns of panels in Fig.\ \ref{fig:class2}, 
the QP-HOMO states are primarily located on the A nucleobases. 
In GpC and CpG, characterized in the last two columns of panels, we see the same pattern, with the QP-HOMO state primarily located on G. 
For ApT, we find a dipole moment close to zero, while TpA and CpG shows small dipoles.
Only the CpG system has a large dipole (Table \ref{tab:OpticalPropPredict}).

Interestingly, in both ApT and TpA the QP-LUMO lobe is shared between A and T, with higher concentration on T. 
Similarly, in both GpC and CpG the QP-LUMOs have weight at both ends, and it is only for the CpG system that the QP-LUMO lobe is concentrated to be located outside the two G nitrogen bases.  
The dipoles of Class-2 WCds are likely aiding in setting the small QP-LUMO differences between the ApT and TpA cases as well as between the GpC and CpG cases, bottom row of panels in Fig.\ \ref{fig:class2}. 
However, a general trend of double-end lobe positions for the QP-LUMOs cannot be a reflection of trapping by the WCd dipoles (as suggested by the analysis of QP-LUMOs in individual WC pairs, above). 
Nevertheless, we can interpret the set of QP-LUMOs of Class-2 WCds as being results of electrostatic trapping.

Figure \ref{fig:class2dipoles} clarifies at the same time (along with Table \ref{tab:OpticalPropPredict}) that quadrupole moments play a significant role in setting the electrostatics of these WCds.  
These WCd are symmetric, and that impacts the trapping of the QP-LUMO states. We note that the resulting quadrupole moments 
are large in Class-2 WCds.
We first reconsider the ApT system (upper left panel), noting that the 
vanishing dipole follows only because the projection of the constituent-dipoles (onto the plane of the WC-pairs) are almost exactly opposite, when viewed in the plane of the WC pairs. These constituent dipoles are still residing on two different WC-pairs and can thus produce a reasonably large quadrupole moment, Table \ref{tab:OpticalPropPredict}. 
With an electrostatic-trapping mechanism defined by the field of the ApT quadrupole, we should expect the trapping to be available at both ends of the stack. This is indeed also what AHBR-mRSH* predicts for the ApT QP-LUMO.

Considering next the TpA and CpG cases, analyzed in the bottom two panels of Fig.\ \ref{fig:class2dipoles}, 
we find a similar mechanism for (now partial) cancellation in the WCd dipole. 
However, Eqs.~(\ref{eqA:qxz}) and (\ref{eqA:qyz}) of Appendix A give a clear argument for the emergence of 
(instead) a large quadrupole moment. 
The partial cancellations in the dipole are causing, instead, strong quadrupole moments of these WCds,  significantly so for the case of CpG.
Accordingly, we can again expect that electrostatics leads to the formation of QP-LUMOs lobes at both ends, 
as we find in our AHBR-mRSH* characterizations.

Turning to the GpC system (upper right panel of Fig.\ \ref{fig:class2dipoles}), we note that the relative twist of the constituent WC-pairs is such that some component of the dipoles survives to point towards the minor groove. 
This dipole is not pointing to the ends where the QP-LUMO lobes of the GpC are actually found. 
However, GpC is a WCd that has one of the two largest quadrupole moments, Table \ref{tab:OpticalPropPredict} and again it is reasonable to expect that electrostatics
(dipoles and quadrupole together) produce a double-lobe character for trapping of an 
electron approaching in a sudden EA event, as predicted  in panel 
k) of Fig.\ \ref{fig:class2}.

Finally, Fig.\ \ref{fig:class3} presents our characterizations of the frontier QPs of WCds Class 3, i.e., for ApG, TpC, ApC, and TpG.
In all four WCd, we find that the QP-HOMO state is mainly located on G, with a much smaller contribution from the nucleobase with which G is stacked (A in ApG and TpC, T in ApC and TpG). 
The QP-LUMO is located at the end of the WCd that does not contain G, and 
the dipole vectors, shown in the middle row of panels, are now pointing toward the QP-LUMO states, as in the class-1 WCds.
We observe significant spatial distribution of the large QP-LUMO lobe in all four cases, extending also between the two WC pairs.

In summary, Figs.~\ref{fig:AT_CG_ahbrmrsh}-\ref{fig:class3} reveal an overall trend: 
The QP-HOMO and QP-LUMO levels are generally pushed onto different molecules of both individual WC-pairs and when these are stacked. Moreover, the QP-HOMO-QP-LUMO wavefunction overlap is further decreased because the QP-LUMO states are always pushed to an outside location 
(like for the individual nucleobases, Figure \ref{fig:nuc_ahbrmrsh}), away from where the hydrogen bonding occurs in the WC pairing. 
The QP-LUMO states are often located on the side indicated by the direction of the computed polarization vectors (arrows in middle panels). Exceptions arise only for the Class-2 WCds
where the dipoles of the constituent WC pairs 
are compensated for symmetry reasons.
There we can instead interpret the set of resulting double-end QP LUMOs as set by electrostatic trapping. 
The implication is that we can generally interpret (sudden) EA events as occurring away from atoms, and not in VBSs. 
This follows because the QP-LUMO results are descriptions of virtual excitations corresponding to these 
events \cite{hedin1965,AuJoWi00,nguyen2016,reining2018}.

\subsection{Quantitative characterization}

We supplement the qualitative analysis presented in the previous sections with a more quantitative one. 
 
It is important to note that our AHBR-mRSH* functional is designed with an asymptotic Fock exchange that reflects the molecular nature of the investigated nitrogen-base systems.
Also, the procedure that we use for the setting the OT AHBR-mRSH*
requires nothing more than self-consistency in calculations that we obtain by use of AHBR-mRSH*  itself as a hybrid-functional DFT. Our method is not adjusted ad-hoc to match QP-HOMO values or QP-LUMO values, as is further explained Ref.~\onlinecite{schroder2025}.

\begin{figure}[tb]
    \centering
        \includegraphics[width=8.5cm]{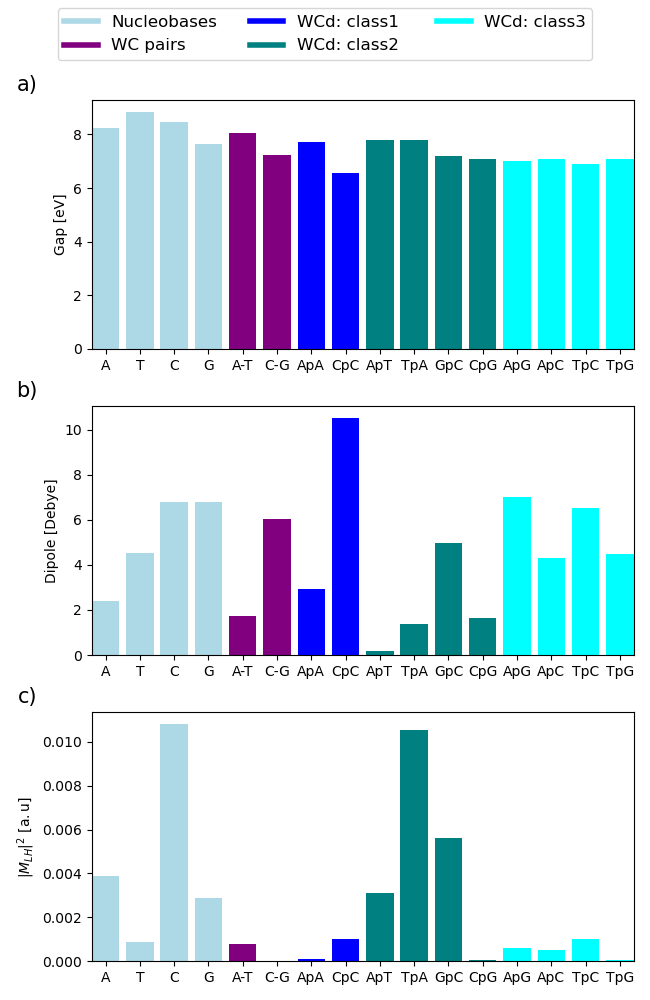} 
    \caption{(a) Energy gap $\Delta_{\rm LH}$, (b) length of dipole moment $d$, and (c) transition dipole moments 
    $|M_{\rm LH}|^2$ of nucleobases, Watson-Crick (WC) pairs and WC pair dimers (WCd), computed in AHBR-mRSH*.}
    \label{fig:gap_dipol_matrix}
\end{figure}

Fig.\ \ref{fig:gap_dipol_matrix} shows three bar charts, with horizontal axis representing the systems under study. 
In panel a) we plot the values of the QP-HOMO-LUMO gap $\Delta_{\rm LH}$ from Table \ref{tab:OpticalPropPredict}. 
The largest gaps are found in the nucleobases, and the smallest gaps are observed in the Class 3 WC dimers, although the differences are small. 
We note (Table \ref{tab:OpticalPropPredict}) that whenever two components are brought together (two nucleobases into a WC pair, two WC pairs into a WCd) the resulting structure has lower value of $\Delta_{\rm LH}$ than each of the components.
We interpret these changes as resulting from hydrogen bonds in the WC pairs and the vdW binding in the WCds.

Fig.\ \ref{fig:gap_dipol_matrix}(b) and Table \ref{tab:OpticalPropPredict} show the magnitude of the dipole moments $d$ of the various systems, as obtained from our calculations. 
The values obtained for the nucleobases are  close to those found in previous studies \cite{Vovusha2018}. 
It is noteworthy that (opposite to the $\Delta_{\rm LH}$ behavior) the highest values of $d$ correspond to C and G, which each have only one oxygen atom that pulls at the electrons. 
For WC pairs, there is a clear decrease in dipole moment magnitude compared to the individual nucleobases, especially in A-T pairs. Again, we interpret this decrease as a consequence of redistribution of charge that arises with the hydrogen bonding.

The dipoles in the WCds are approximately the sum of dipoles of the WC pairs, as discussed in the Appendix. 
For Class 1 WCds, where the two WC pairs are the same (but at a relative twist), the sum of their dipoles is almost double the size. This is reflected in the values reported in panel b) of Fig.~\ref{fig:gap_dipol_matrix}.
In the Class 2 WCd, the WC pairs are again the same, but one is flipped, as illustrated in Figure \ref{fig:class2dipoles}. 
The dipoles of these WC pair partially cancel each other, and the resulting WCd dipoles are here smaller than that of each of its constituent WC pairs.
For the WCds of Class 3 all four nucleobases of the dimers are different. The value of $d$ then depends on whether the WC pair dipoles are (roughly) parallel, antiparallel, or just a mix, as illustrated in the middle panels of Fig.\ \ref{fig:class3}. 

Related to the (magnitude of the) dipole moment, we also determine and discuss the squared absolute value of the transition dipole moments 
$|M_{\rm LH}|^2$.
Panel c) of Fig.\ \ref{fig:gap_dipol_matrix} and Table \ref{tab:OpticalPropPredict} presents an overview of the numerical values. Again, this quantity is associated with the strength of interaction with an electromagnetic wave. We observe that among the nucleobases, C has the largest magnitude on the matrix element, while the smallest is that of T. 
As seen in Figure \ref{fig:nuc_ahbrmrsh} top and bottom, the spatial overlaps of the QP-HOMO (top) and QP-LUMO (bottom) states are appreciable, and so are the values of $|M_{\rm LH}|^2$. However, the WC pairs exhibit a drastic reduction in these values compared to the individual nucleobases because of spatial separation of the QP-HOMO and QP-LUMO. 
The A-T pair is less reduced than the C-G pair, which has an almost zero transition moment.
This is because the QP-LUMO state in A-T, while being mostly located on the edge of T, also has a small presence on the edge of A, panel e) of Fig.~\ref{fig:AT_CG_ahbrmrsh}.

Among the WCds, those of Class 2 are distinct because both the QP-HOMO and QP-LUMO states are distributed on both ends of the WCd, Fig.~\ref{fig:class2}. This again gives an
appreciable overlap between the states and thus the value of $|M_{\rm LH}|^2$, as seen in panel c) of Fig.~\ref{fig:gap_dipol_matrix}(c). An exception is 
CpG  where the QP-LUMO is mostly distributed on G, thus $|M_{\rm LH}|^2$ nearly
vanishes.
In the other types of WCds, Fig.~\ref{fig:class1} and \ref{fig:class3}, the QP-HOMO and QP-LUMO are clearly separated, 
distributed on opposite ends of the WCd, making the QP overlap small.

Table~\ref{tab:NucPlusWCpairs} presents our results for the frontier-QP energies of the investigated nucleobase systems. For the four isolated nucleobases and for the two WC pairs,
we include comparisons with experimental data
and other accurate theory results for QP energies in these type of systems \cite{ DoYoVo78,DeAbSc96,roca2009,HeLyCl96,TrScKo06,tripathi2019}. The theory-experiment comparison, for the last two occupied QP states (QP-HOMO-1 and QP-HOMO) and for the first two unoccupied or virtual QP states (QP-LUMO and QP-LUMO+1), is partly repeated from the AHBR-mRSH* launch study \cite{schroder2025} but also expanded to cover nitrogen systems up the WCds.

Table \ref{tab:NucPlusWCpairs} shows QP predictions by AHBR-mRSH* for isolated nitrogen bases and for WC pairs that are all in good agreement with previous QP characterizations. 
For the nucleobases, we find occupied QP energies that are generally close to the experimental values, with the difference consistently being an underestimate compared to the experimental results. On average, the deviation for QP-HOMO is $\sim 0.2$ eV compared to the experimental values.
Similarly, for the QP-LUMO values of the nucleobases, we observe that A does not exhibit a dipole-trapped bound state, as it has a positive eigenvalue. This is correctly captured in our AHBR-mRSH* characterization. In the case of T and G, the functional accurately captures the nature of the dipole-trapped states as very weakly bound. For G AHBR-mRSH* also predicts a dipole-trapped QP-LUMO but no experimental data is available for comparison.

For the WC pairs A-T and C-G, we compare our results with predictions of vertical electron affinities that have been computed using the 
high-precision EA-EOM-DLPNO-CCSD theory \cite{tripathi2019}. 
In general, the differences between the EA-EOM-DLPNO-CCSD and AHBR-mRSH* results for WC pairs are on the order of meVs. 
For the A-T pair, we observe that in our calculations the electron is not bound, while it is very weakly bound according to EA-EOM-DLPNO-CCSD. 
In the C-G pair, the extra electron of a sudden EA event (corresponding to the QP-LUMO) is slightly bound, with both methods showing consistent results.
In summary, the ABHR-mRSH* works as a descriptor of nitrogen-base QPs, as documented so far up to WC pairs.

Moving next to what we may call AHBR-mRSH* \textit{QP predictions},
we also discuss the  nature of the QP LUMOs,
following Refs.\ \onlinecite{nguyen2016,schroder2025}.  Experimental observations and theory results shows, where available, that the actual QP-LUMO states in the nitrogen base systems has a dipole trapped character \cite{nguyen2016}. They also show that at least some of the following (higher-energy) unoccupied QP should have a VBS character. It is not easy for a theory method to be right about both of these qualitative results at the same time. It is also not easy to summarize the success that a QP theory (like AHBR-mRSH*) has at meeting this challenge.

Still, in Table \ref{tab:NucPlusWCpairs} we implement the idea \cite{nguyen2016} 
for a quantitative assessment. That is we a)  nominally label the first unoccupied QP with a dipole-trapped character as `QP-LUMO' and the other low-energy QP-LUMO (ideally an unoccupied VBS) as the `QP-LUMO+', AND b) test
method quality in terms of the predicted energy ordering of those two states. We supplement this assessment, 
in Table~\ref{tab:NucPlusWCpairs}, with a discussion of whether AHBR-mRSH* predict these states as bound or as existing instead as unbound, i.e., predicted as resonances.

Table \ref{tab:NucPlusWCpairs} shows that for all investigated systems AHBR-mRSH* delivers what either is or (for WCds) what is expected to be qualitatively correct energy orderings. As also shown in previous figures mapping the QP-LUMO nature, the AHBR-mRSH* does provide a correct placement of
electrostatically trapped states as the 
lowest-energy empty QPs, i.e., as what we have now nominally labeled as QP-LUMOs.

We find a slightly bound QP-LUMO+1 state in the T nucleobase (while 
AHBR-mRSH* predicts positive QP-LUMO+1 energies for the other bases). 
We also predict a slightly bound QP-LUMO+1 state for the C-G WC pair and for the CpC and GpC WCd. 
Interestingly, the spatial distribution of the LUMO+1 state on T is concentrated \textit{on} T and resembles the KS-state distribution (shown in Fig.~\ref{fig:GT_b86r}), which itself mimics a VBS.
The nature of the experiments for these electron-attachment events is such that they involve electron transfer from photo-excitation of surrounding water molecules \cite{nguyen2011}. 
It is therefore motivated to assume that adiabatic effects, i.e., impact of atomic relaxations play are role. The energy ordering with the dipole- and multipole trapping occurring before a VBS-type LUMO has also been found in other theory studies \cite{liu2020}.
However, unlike in those theory studies, for our finding of VBS at LUMO+1 energy positions, we did not allow any further atomic relaxation in our AHBR-mRSH* predictions. With AHBR-mRSH* we find the VBS character as occurring directly for the QP-LUMO+ states in several of the nitrogen-base systems, in agreement with experiments \cite{svozil2005,DeAbSc96,schroder2025}.

Considering the nature of the QP-LUMO+ states in WC pairs, we find that it is only for C-G that AHBR-mRSH* stabilizes both QP-LUMO and QP-LUMO+1 states. This is an effect that we interpret as correlated with 
the fact that this WC pair has a high  dipole moment. The spatial distribution of its QP-LUMO+1 state is again on the atoms, as for the T nucleobase. 

Finally, we consider the set of broader AHBR-mRSH* \textit{QP predictions} for the set WCds, lower sections of Table~\ref{tab:NucPlusWCpairs}. Here we observe that AHBR-mRSH* do not generally stabilize the QP-LUMO+1 states of these WCds as bound. 
Still, two exceptions are CpC and GpC, and these are also the systems that have the largest net dipoles.
This systematic variation in AHBR-mRSH* results is reasonable but also invites further testing.

\section{Conclusion}

This study demonstrates that the optimally tuned range-separated hybrid functional AHBR-mRSH* provides a computationally efficient 
characterization of the electron-attached and ionized states of key DNA components. 
Our approach reveals insights into dipole-bound states and QP properties, as well as descriptors of possible optical transitions  
at the gap, i.e., between the QP-HOMO 
and QP-LUMO states.
Also, 
by systematically analyzing nucleobases, WC pairs, and WC-pair dimers, we arrive at a classification based on frontier states, 
dipole and multipole moments, as well as transition dipole moments. This is a framework for seeking  
a better understanding of the photophysical behavior of DNA.

With this paper, we hope to help invite more theory or measurement works on the QP nature in DNA subsystems, for example, the WCds. 
We find that AHBR-mRSH* delivers accurate QP predictions in cases where we can compare with existing data. Broader validation would further heighten the trust that we have in AHBR-mRSH* as a general-purpose QP predictor  for (small) nitrogen-based systems. 
Conversely, documentation of deviations from QP predictions by AHBR-mRSH* may instead drive us to improved methods for capturing the nature of DNA QPs.

As a first overall observation on our QP predictions, we stress the qualitative differences arising for the QP and KP LUMO states but 
not for the HOMO states.
As a second overall observation, we note that the QP-LUMOs correspond to dipole- or multipole-trapped anions as generated by electron attachments followed by photon emission. 
In cases where there is a large dipole, the QP-LUMO states sit roughly at the end of that dipole; In other cases the quadrupole is important, but the dipoles located by each of the WC pairs are still 
driving the QP-LUMO states to set away from the (central) hydrogen-bonding region of the WC pairs.

A third overall observation is simply that our finding of a systematic multipole trapping means that we can have some trust in our QP-LUMOs as a characterization of charged, 
so-called vertical excitations.  
With B86R and AHBR-mRSH* studies we inherit the Born-Oppenheimer approximation. As a consequence, we here model the electron attachment while ignoring the atomic relaxations,
as is often done in MBPT studies. 
However, with a valence-trapped LUMO (as found in the KS-DFT description, Fig.\ \ref{fig:GT_b86r}), the added 
charge sits among the original electron density so that we should then expect structural distortions to follow, at least on a pico-second time scale. 
In contrast, with the present AHBR-mRSH* QP-LUMO predictions, we find a significantly larger spatial separation and can expect significantly less 
atomic-distortion effects to cause less adjustments for our QP-LUMO predictions by atomic-relaxation effects. 

Finally, we include a brief discussion of  possible implications
existing for EA-induced DNA damage \cite{zheng2006}.
We are here predicting that electrostatics 
of the WC pairs and WCds tends to drive the extra electron in a sudden EA event out towards the DNA back bone, on either one or both sides. 
This finding is provided in studies that do not include the DNA backbone structure. 
However, there are experimental observations that electron capture at any given base system determines whether 
there are ensuing ruptures of the backbone exactly outside the specific base, Ref.\ \onlinecite{zheng2006}.
This observation is 
not inconsistent with the electrostatic trapping mechanisms that we have here sought to characterize.

\section{Acknowledgment}
The present work is supported by the Swedish Research Council (VR) through Grants No.\ 2020-04997 and 2022-03277, 
and by the Chalmers Area of Advance (AoA) Nano
and Chalmers AoA Production.
The computations were performed using computational and storage resources at 
Chalmers Centre for Computational Science and Engineering (C3SE), 
and with computer and storage allocations from the 
National Academic Infrastructure for Supercomputing in Sweden (NAISS), under contracts
NAISS2023/3-22, 
NAISS2023/6-306, 
NAISS2024/3-16, 
and NAISS2024/6-432. 

\appendix
\section{Estimates of dipoles and quadrupoles of WC dimers}
From the (DFT-calculated) dipoles $\vec{d}$ of the WC pairs  
we estimate the dipoles and quadrupoles of the WC dimers.
In a coordinate system that has origin in the DNA center midway 
between the two WC pairs, the $z$-axis along the DNA center, 
and further oriented such that the two WC pairs are at 
$-36^\circ/2$ (lower pair) and $36^\circ/2$ (upper pair) with the $y$-axis, the WC-pair 
dipole end point positions can be written
\begin{eqnarray}
\vec{r}_{\pm 1} & = & (\pm x_1, \pm y_1, -\delta /2) \\
\vec{r}_{\pm 2} & = & (\pm x_2, \pm y_2, \delta /2 )
\end{eqnarray}
where $\delta=3.3$ {\AA} is the distance between the WC pairs, and the charge at each end is $\pm q_1$ and $\pm q_2$
with the positive $q_1$ and $q_2$ at the tip of the dipoles. 
This all assumes that the dipoles sit at the DNA center. 
In practice, for our WC pairs, the dipoles of A-T and C-G sit just 0.4 {\AA} and 0.6 {\AA} away from the DNA center, 
making the assumption valid.

The components of our approximation for the WC dimer dipole, $\vec{d}_\Sigma$, are estimated from adding the four assumed charges $\pm q_1$ and $\pm q_2$ of the 
two WC pair dipoles at their positions
\begin{equation}
d_{\Sigma,i} = \sum_{l=-2,-1,1,2} q_l r_{il} \, .
\end{equation}
Inserting the charges and positions
also leads us to an approximation for the 
component of the quadrupole in the traceless form
\begin{equation}
Q_{ij} = \frac{1}{2} \sum_{l=-2,-1,1,2} q_l (3  r_{il}r_{jl} - \delta_{ij}\sum_{k=x,y,z} r_{kl}^2 ) \, . 
\label{eq:Qij}
\end{equation}
Here $l$ cycles through the end points of the two dipoles $|l|= 1$ and 2, 
$q_l= -q_{-l}$ is the charge of dipole $|l|$, and
$r_{il}$, $r_{jl}$, and $r_{kl}$ are the $i$'th, $j$'th, and $k$'th coordinates 
($i,j,k=x$, $y$, $z$) for WC-pair dipole $|l|$ of the start ($l<0$) and end 
($l>0$) of the dipole vector. 

In effect, the estimated WCd dipole is simply 
$\vec{d}_{\Sigma} = \vec{d}_1 + \vec{d}_2$
and we find, by insertion into Eq.\ (\ref{eq:Qij}), 
that all but four entries in the quadrupole matrix $\textbf{Q}$ vanish with our choice of coordinate system. The nonzero
components of the quadrupole moment are
\begin{eqnarray}
Q_{xz} & = & Q_{zx} = 3\delta (q_2x_2-q_1x_1)/2 = \frac{3}{4} \delta \Delta d_x
\, ,
\label{eqA:qxz}\\
Q_{yz} & = & Q_{zy} = 3\delta (q_2y_2-q_1y_1)/2 = \frac{3}{4} \delta \Delta d_y \, .
\label{eqA:qyz}
\end{eqnarray}
Here $\Delta d_i$ denotes the $i$'th component of the difference in the two WC pair dipoles,
$\Delta \vec{d} = \vec{d}_2 - \vec{d}_1$.
The three eigenvalues of $\mathbf{Q}$ are 0 and $\pm \frac{3}{4} \delta |\Delta \vec{d}|$.

From just two DFT calculations of the dipoles $\vec{d}$, one each for the two WC pairs A-T and C-G 
(or experimental measurements of the same), we can obtain estimates of the dipole vector and quadrupole matrix for
all ten WC dimers,
by suitable rotations, and (when relevant) flipping of the dipoles of the two WC pairs. 
In particular, in ApT, TpA, CpG and GpC we have $x_2=-x_1$ and $y_2=y_1$ for symmetry reasons, 
which, in turn, means $Q_{yz}=Q_{zy} = 0$. 

Table \ref{tab:OpticalPropPredict} presents calculated numerical values for the nonzero components of the quadrupole moments, as well as for the largest eigenvalue, for the set of ten WCds.

\bibliography{bib} 
\bibliographystyle{rsc}
\end{document}